\documentclass[12pt,a4paper]{iopart}

\usepackage{graphicx}
\usepackage{subcaption}
\usepackage{aas_macros}
\usepackage[utf8]{inputenc}
\usepackage[T1]{fontenc}
\usepackage{mathrsfs}
\usepackage{latexsym,amsfonts}
\usepackage{bm}
\usepackage[dvipsnames]{xcolor} 
\usepackage[
    colorlinks=true,citecolor=blue,
    linkcolor=blue,
    urlcolor=magenta]{hyperref}

\usepackage{tikz}
\usetikzlibrary{arrows.meta, positioning}

\expandafter\let\csname equation*\endcsname\relax
\expandafter\let\csname endequation*\endcsname\relax

\usepackage{amsmath}
\usepackage{amssymb}
\usepackage[mathlines]{lineno}
\usepackage{algorithm}
\usepackage{algpseudocode}
\usepackage{multicol}

\usepackage{lineno}
\allowdisplaybreaks
\eqnobysec

\def\tdfstat{\texttt{TDFstat}}
\newcommand{\fstat}{{$\mathcal{F}$-statistic}}
\def\fcal{\mathcal{F}} 

\begin{document}

\title[Deterministic denoising of long-duration GW signal candidates with $\alpha$-(de)blending]
{Deterministic denoising of long-duration gravitational-wave signal candidates with $\alpha$-(de)blending}

\author{Przemys{\l}aw Figura$^1$ and Micha{\l} Bejger$^{1,2}$}

\address{$^1$ Nicolaus Copernicus Astronomical Center, Polish Academy of Sciences, 00-716, Warsaw, Poland}
\address{$^2$ INFN Sezione di Ferrara, Via Saragat 1, 44122 Ferrara, Italy}

\begin{abstract}
We present a deterministic machine-learning method for denoising candidate signals from all-sky searches for continuous gravitational waves from rotating neutron stars. Using the time-domain $\fcal$-statistic search pipeline, we post-process the resulting $\fcal(f,\dot{f})$ candidate patterns with iterative $\alpha$-(de)blending, a deterministic diffusion-type generative model implemented as a U-Net, well suited to the non-Gaussian, correlated noise of the $\fcal$-statistic output. Two models, trained on data based on simulated 6- and 12-day time-domain segments with software-injected signals similar to the LVK hardware injections, are tested by comparing deblended images to the library of expected signal patterns via the Structural Similarity Index Measure. In general the method recovers the correct sky-position-dependent pattern for signal-to-noise ratios $\rho \gtrsim 4$, moderately below typical all-sky detection thresholds, with recovery depending strongly on pattern morphology, demonstrating that deterministic diffusion-based denoising may serve as a consistency/veto tool ahead of the semi-coherent coincidence stage. We also discuss limitations and possible improvements of the method. 
\end{abstract}

%\maketitle

\section{Introduction}
\label{sec:intro} 

Gravitational waves (GWs) are perturbations of spacetime, detectable since 2015 \cite{GW150914} with laser interferometers of the LIGO-Virgo-KAGRA (LVK) Collaboration \cite{ALIGO2015,AdV2015,KAGRA2013}. Currently, 390 GW short-duration events, emitted by binary black hole (BH) systems, binary neutron stars (NSs) and NS-BH systems were announced \cite{gwtc5}. Other, still undetected signal types are long-duration, usually called continuous GWs (CWs), originating from rotating, non-axisymmetric NSs (although other astrophysical processes, e.g.~hypothetical boson clouds around BHs or inspiraling light primordial BH systems are also discussed \cite{Haskell2023}). 

Since CWs have smaller GW amplitudes than the already-detected transient signals, but persist for durations comparable with the observing time, one has to analyse long segments of data to detect them. Search strategy depends on the amount of information on the possible CW source: targeted searches towards e.g.~known pulsars assume known location and frequency parameters (see \cite{2025ApJ...983...99A} and references therein for previous searches). It is the most fine-tuned type of search for exactly the chosen parameters in coherent integration data over the longest periods. Directed searches aim at specific sky regions without assuming narrow frequency parameter priors, as in e.g.~searchers towards supernova remnants (see \cite{2026arXiv260325808T} and references therein). All-sky searches are agnostic in terms of sky position and frequency parameters, and therefore computationally-demanding searches for CW sources anywhere in the sky with wide range of possible rotational parameters (see \cite{2026arXiv260314168T,PhysRevD.106.102008} and references therein). Due to the size of the parameter space, CW signals are searched for in data segments based time-domain input much shorter than the observing run, and then coincidence search between signal candidates in different segments is performed (a semi-coherent search). 

In this work we present a CW candidate signal post-processing method, i.e.~recovering CW signal signature (pattern) with machine learning (ML) denoising (specifically, deterministic $\alpha$-(de)blending \cite{alpha-deblending}) of a coherent search data output, to increase the significance of a signal candidate (or falsify/veto it) for the next step of a semi-coherent search, i.e.~the search for coincidences. We use a specific CW search method, namely the time-domain $\fcal$-statistic (\tdfstat) method \cite{1998PhRvD..58f3001J}, initially designed for an all-sky search on narrowband time-domain segments \cite{tdfstat-repo}. In Secs.~\ref{sec:TDFmethod} and~\ref{sec:fstat_pattern} we describe the {\tdfstat} method; the GW data and how it is processed to become the input to the denoising method is described in Sec.~\ref{sec:data}. Section~\ref{sec:MLmethod} describes the ML technique we use (Sec.~\ref{sec:IADB}) and its neural-network (NN) implementation  (Sec.~\ref{sec:nn}), following with results of the analysis in Sec.~\ref{sec:results}. Section~\ref{sec:summary} contains a summary and conclusions.

\section{Time domain F-statistic search} 
\label{sec:fstat}

\subsection{The method} 
\label{sec:TDFmethod}

Input data for the denoising procedure consists of an output produced by one of the standard GW data analysis methods applied to searches for CWs: the {\tdfstat} method \cite{1998PhRvD..58f3001J} (see also \cite{PhysRevD.59.063003,PhysRevD.61.062001,PhysRevD.65.042003,PhysRevD.82.022005} for details and algorithmic improvements). {\tdfstat} provides an optimal (maximum-likelihood) coherent search detection statistic for long-duration, nearly-periodic GW signals, also called quasi-monochromatic because the GW frequency is changing slowly, $f\approx\mathrm{const.}$ A promising source of this kind of signals are rotating, non-axisymmetric NSs; for recent reviews see \cite{Sieniawska2019,Riles2023,Haskell2023,WETTE2023102880}. The $\fcal$-statistic takes into account the phase modulation caused by the Doppler modulation due to the movement of the detector (located on Earth) with respect to the GW source, as well as the GW amplitude modulation due to moving detector's antenna pattern. In addition, signal's phase can be modulated intrinsically due to the change of the GW frequency of the source, which is reflected by non-zero frequency time derivatives $\dot{f}$, $\ddot{f}$, \dots~(so-called spindowns).  

The detector's data is a time series $x(t) = n(t) + h(t)$, with $n(t)$ denoting detector's noise at time $t$, and $h(t)$ the GW signal amplitude. The latter is described by
\begin{eqnarray}
\label{TotalH}
    \begin{aligned}
        h(t) =h_0 \Bigl(F_+(t,\alpha_s,\delta,\psi)\frac{1+\cos^2{\iota}}{2} \cos{\phi(t)} \Bigr. + \Bigl. F_{\times}(t,\alpha_s,\delta,\psi) \cos{\iota} \sin{\phi(t)} \Bigr) \, ,
    \label{eq:ht}
    \end{aligned}
\end{eqnarray}
where $h_0$ is the intrinsic GW amplitude of the signal, assumed here to be constant, $F_+$ and $F_{\times}$ are detector's antenna pattern functions, $\delta$ is the source declination, $\alpha_s$ its right ascension, $\psi$ is the signal polarisation angle, $\iota$ is an angle between total angular momentum of the source and the direction from the source to the detector, and $\phi(t)$ is the signal's phase. One can approximate the phase by a Taylor expansion, and for slowly varying GW frequency, the following approximation with frequency $f$ and frequency time derivative $\dot{f}$ is sufficient: 
\begin{eqnarray}
    \begin{aligned}
        \phi(t) = \phi_0 + 2\pi f \Bigl( t + \frac{{\bf n} \cdot {\bf r}_d(t)}{c} \Bigr) + 2\pi \dot{f} \left (t + \frac{{\bf n} \cdot {\bf r}_d(t)}{c} \right)^2 \, ,
    \label{eq:phaseevo}    
    \end{aligned}
\end{eqnarray}
where $\phi_0$ is an initial phase. The LVK GW detectors are located on the rotating Earth orbiting the Sun. To take these modulations into account, the data is represented in the Solar System Barycenter (SSB) coordinate frame, where ${\bf r}_d(t)$ is a vector joining the SSB with the detector, and ${\bf n}$ is a unit vector pointing from SSB to the source. If the GW source is a triaxial rotating NS spinning at the frequency $f/2$ then its GW amplitude $h_0$ is 
\begin{eqnarray}
    \begin{aligned}
        h_0 &= \frac{4\pi^2G}{c^4} \frac{\epsilon I_{zz} f^2}{d}  \\
        & \approx 1.06\times10^{-26} \left(\frac{\epsilon}{10^{-6}}\right)\left(\frac{I_{zz}}{10^{38}\>\rm{kg\ m}^2}\right)\left(\frac{f}{100\>{\rm  Hz}}\right)^2
\left(\frac{1\>{\rm kpc}}{d}\right) \, , 
    \label{eqn:hexpected}
    \end{aligned} 
\end{eqnarray}
where $d$ is the distance to the source, $I_{xx}$, $I_{yy}$, and $I_{zz}$ are the principal moments of the star inertia, and $\epsilon=(I_{xx}-I_{yy})/I_{zz}$ is the ellipticity of the NS. 

A complete derivation of the {\fstat} method is provided in \cite{1998PhRvD..58f3001J}. Here we note that the {\fstat} is, in its core, a matched filter method \cite{1057571}, designed to maximise its response in the presence of a signal $h(t)$ (Eq.~\ref{eq:ht}), which may be expressed as a linear combination of four known time-dependent basis waveforms $h_i$:
\begin{equation}
   h(t;\lambda,A(\xi)) = \sum_{i=1}^{4} A_i(\xi) h_i(t;\lambda), 
   \label{eq:ht_tdfstat}
\end{equation}
where amplitude parameters $A_i$ are known functions of $\xi$, denoting the intrinsic amplitude $h_0$, inclination $\iota$, polarization $\psi$ and initial phase $\phi_0$. The method amounts to analytical maximization of the matched-filter likelihood over parameters $\xi = \{h_0,\iota,\psi,\phi_0\}$ through amplitudes  $A_i(\xi)$ \cite{1998PhRvD..58f3001J}. The remaining unknown parameters: frequency $f$, spindown $\dot{f}$ and sky position $\alpha_s$ and $\delta$, are denoted by 
$\lambda = \{f,\dot{f},\alpha_s,\delta\}$. They are used as search parameters to obtain the final matched-filter {\fstat} value, which is related to the signal-to-noise ratio (SNR) of the signal (see~\cite{1998PhRvD..58f3001J}, Sec.~III for details).

When no GW signal is present in the data, the $2\fcal$ values follow a standard (centrality 0) $\chi^2$ distribution with 4 degrees of freedom, $\chi^2_4$. However, when a signal with an SNR $\rho$ is present, $\fcal$ follows a non-central $\chi^2_4(\rho^2)$ distribution, with the non-centrality parameter $\rho^2$ related to the expected value of $2\fcal$ as $\rho^2 = 2\fcal-4$. Furthermore, the angle-averaged value of $\rho^2$ depends on the signal duration time $T_{s}$ and the signal's intrinsic amplitude $h_0$ as $\langle\rho^2\rangle_{\alpha_s,\delta,\psi,\iota} = \frac{4}{25}h_0^2{T_s/ \tilde S_f}$, where $\tilde S_f$ denotes the power spectral density (PSD) of the detector's noise at frequency $f$ (see Sect.~IIIC in \cite{1998PhRvD..58f3001J}).

In practical terms, search for CW signals with {\em a priori} unknown parameters $\lambda = \{f,\dot{f},\alpha_s,\delta\}$, also referred as an ``agnostic'', ``blind'' or ``all-sky'' search,  amounts to looking for the maximum of $\fcal$ as a function of $\lambda$.  Specifically, we study the output of the all-sky {\fstat} search method: the {\tdfstat} search pipeline \cite{tdfstat-repo,PhysRevD.82.022005}, used routinely on the LIGO and Virgo detectors' data \cite{Aasi_2014,PhysRevD.96.062002,PhysRevD.97.102003,PhysRevD.100.024004,PhysRevD.106.102008}. Although a given CW signal has a well-defined $f$ and $\dot{f}$ values, the output of the {\fstat} is distributed across a range of $f$ and $\dot{f}$ values. Additionally, resulting $\fcal(f,\dot{f})$ distribution depends on signal's parameters: sky position, and additional source orientation angles. 

\begin{figure} 
    \includegraphics[width=0.48\textwidth]{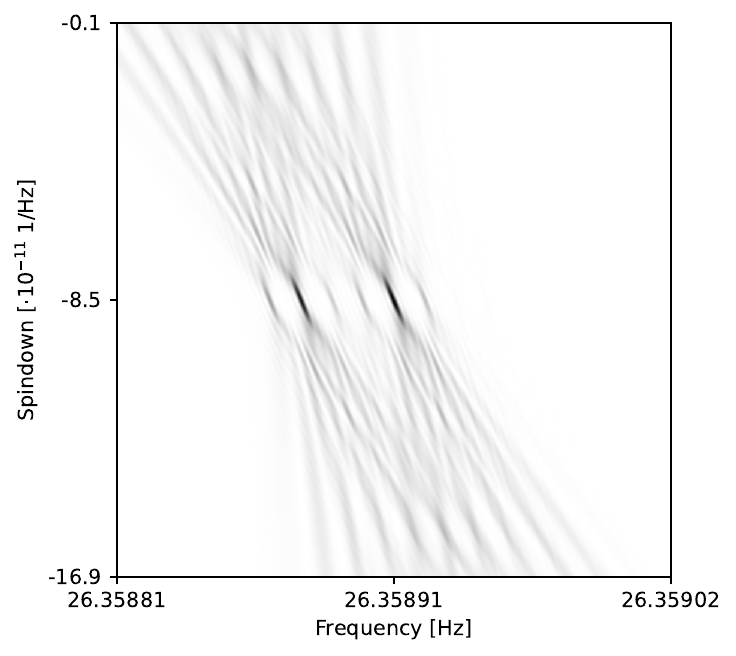}
    \hfill
    \includegraphics[width=0.47\textwidth]{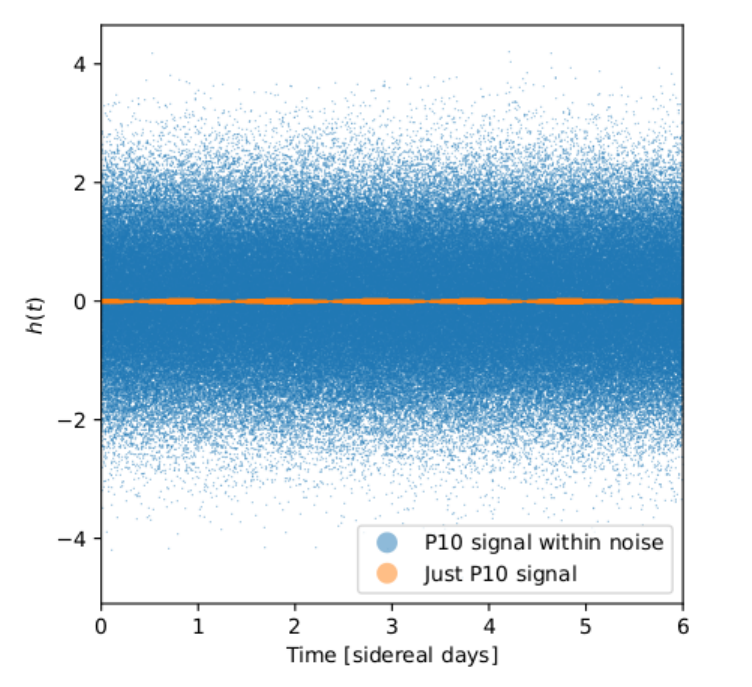}
    \hfill\\
    \includegraphics[width=0.49\textwidth]{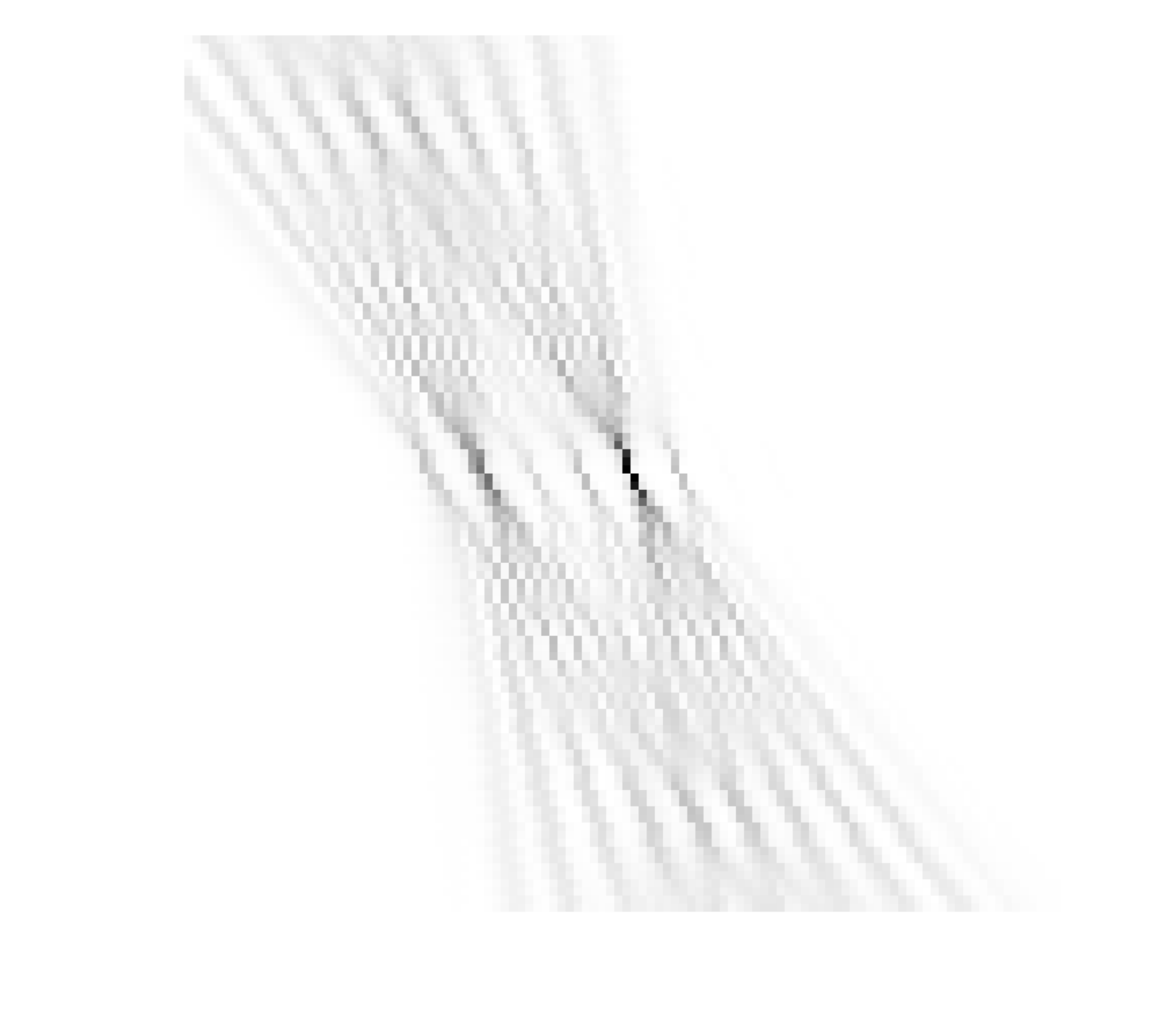}
    \hfill
    \includegraphics[width=0.49\textwidth]{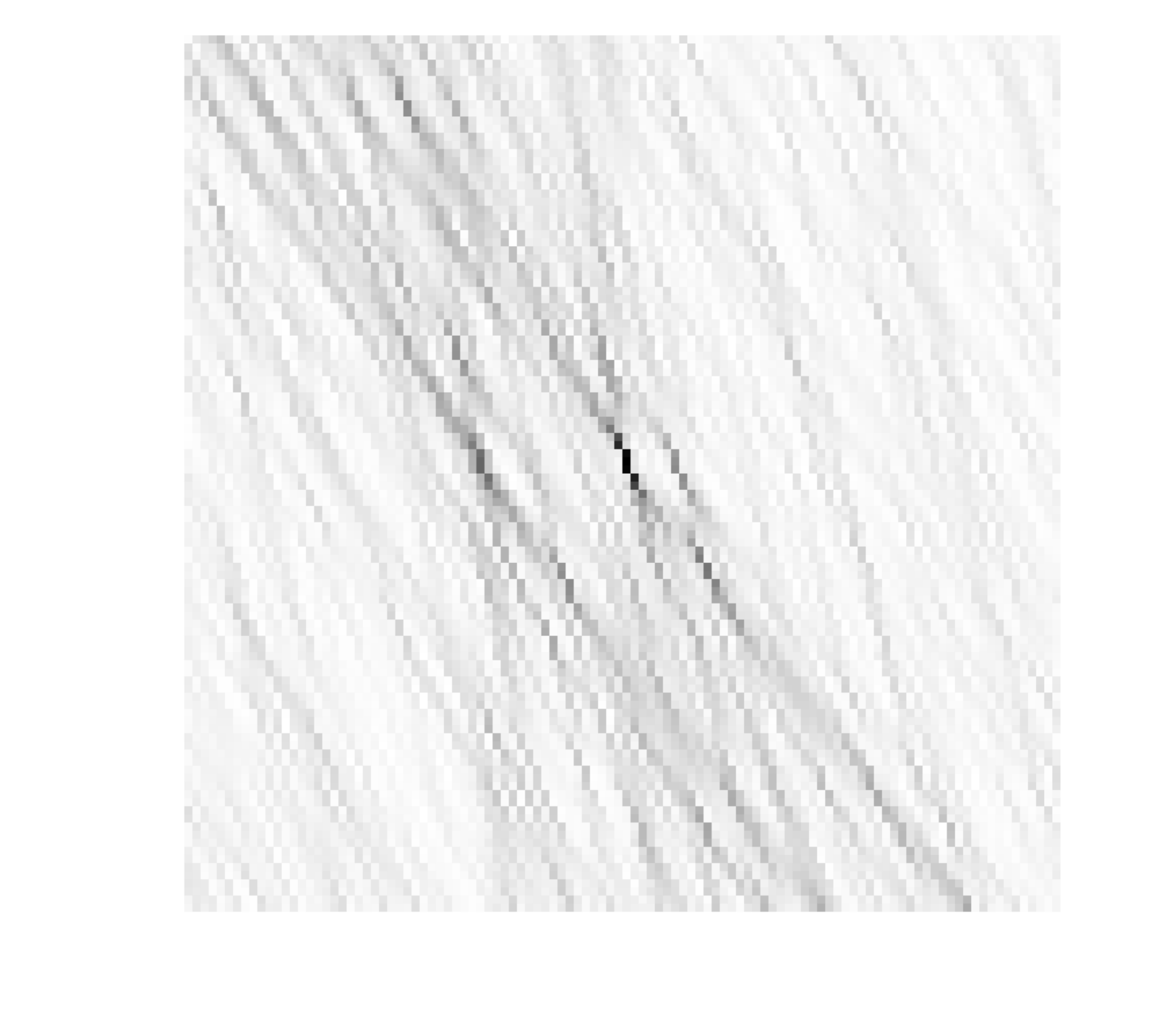}
    \caption{$\fcal(f,\dot{f})$ pattern for a P10 SI, and its time-domain input data. Top left: $\fcal(f,\dot{f})$ pattern in physical units; the maximum of $\fcal$ is located in the center of the panel, coincident with the true $f$ and $\dot{f}$ values of the injection. Top right: 6-days long time series $h(t)$ Gaussian data with CW signal marked in orange (with sidereal day modulations visible) and Gaussian noise in blue; the amplitude in the {\tdfstat} code units. Bottom left: Signal-only $\fcal(f,\dot{f})$ pattern from {\tdfstat} all-sky pipeline for the above time domain input data, in the ``grid'' units (corresponding to orange data in top right). Bottom right: $\fcal(f,\dot{f})$ pattern for signal immersed in noise (corresponding to blue data in top right).} 
    \label{fig:fstat_example}
\end{figure}

In the following we simulate a follow up analysis of {\tdfstat} search signal candidates, consisting of distributions of $\fcal(\lambda)$ values, to check if they are consistent with our CW signal model prediction. Input data to the denoising algorithm (see Sec.~\ref{sec:MLmethod}) consists of 2-dimensional distributions of $\fcal(f,\dot{f})$ for known sky position values $\alpha_s$ and $\delta$, using fixed-width template bank (``grid'') spacing for $f$ and $\dot{f}$, where the $f$--$\dot{f}$ grid is provided by a template bank \cite{2015CQGra..32n5014P,Pisarski_2023}. An example of time domain CW signal data, and resulting {\fstat} distribution of $\fcal(f,\dot{f})$, representing a {\fstat} matched filter output of a CW signal is shown in Fig.~\ref{fig:fstat_example}; see Sec.~\ref{sec:data} for more details.

Figure~\ref{fig:fstat_example} shows a CW signal software injection (SI), in particular SI P10 (see Tab.~\ref{tab:pulsars} and Sec.~\ref{sec:data} for details) added to $T_s = 6$~days segment; sidereal amplitude modulations are visible in time series (top right panel), as well as multiple peaks of the $\fcal(f,\dot{f})$ distribution on {\fstat} image (other panels). For 6-day time domain data, local maxima in the $\fcal(f,\dot{f})$ distribution are separated by 6 pixels, whereas each pixel width in frequency is equivalent to Earth's sidereal rotation frequency, $1.16\times 10^{-5}\,\mathrm{Hz}$. The dependence of the {\fstat} pattern on sky position and other parameters is discussed in more detail in the following section.     

\subsection{F-statistic pattern dependence on source parameters and sky position} 
\label{sec:fstat_pattern} 

As shown in Eqs.~\ref{eq:ht} and \ref{eq:phaseevo}, for a given frequency $f$ and spindown $\dot{f}$, the $\fcal$-statistic depends on signal's intrinsic GW amplitude $h_0$, sky position $\alpha_s$ and $\delta$, as well as angles $\iota$, $\psi$, and $\phi_0$. However, the $\fcal$-statistic pattern in frequency-spindown plane for data segments duration $T_s$ of multiple sidereal days are subject to several simplifications, which are discussed in detail in \ref{sec:fstat_degeneracy}: the pattern does not depend on arbitrary starting phase $\phi_0$, nor on $\alpha_s$ due to averaging effect of Earth's rotation for several days long time segments, and in general $\fcal(\delta, \iota, \psi) = \fcal(-\delta, \iota+\pi, -\psi)$.

\begin{figure}
    \centering
    \makebox[0.8em][c]{\rotatebox{90}{\scriptsize 
    \hspace{1em} $\iota = 0^{\circ}$}}
    \hfill
    \includegraphics[width=0.14\textwidth]{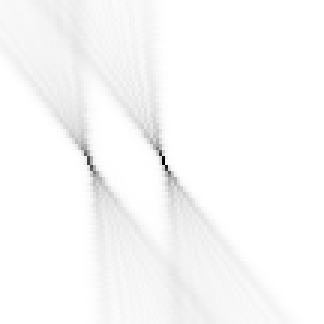}
    \hfill
    \includegraphics[width=0.14\textwidth]{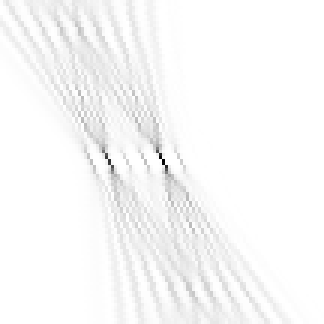}
    \hfill
    \includegraphics[width=0.14\textwidth]{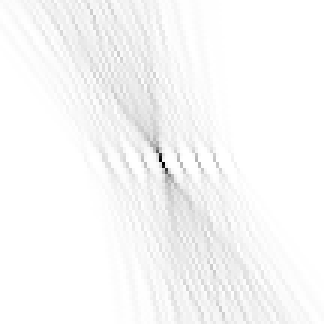}
    \hfill
    \includegraphics[width=0.14\textwidth]{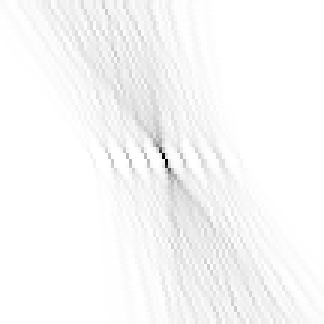}
    \hfill
    \includegraphics[width=0.14\textwidth]{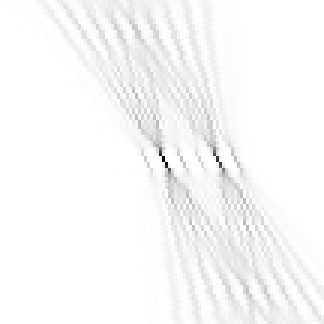}
    \hfill
    \includegraphics[width=0.14\textwidth]{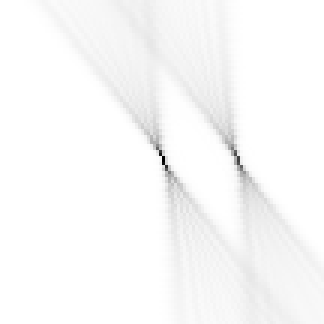} \\
    
    \makebox[0.8em][c]{\rotatebox{90}{\scriptsize \hspace{1em} $\iota = 180^{\circ}$}}
    \hfill
    \includegraphics[width=0.14\textwidth]{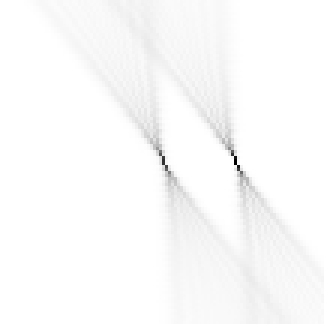}
    \hfill
    \includegraphics[width=0.14\textwidth]{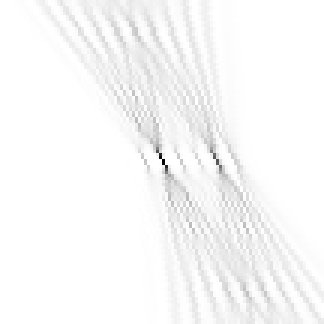}
    \hfill
    \includegraphics[width=0.14\textwidth]{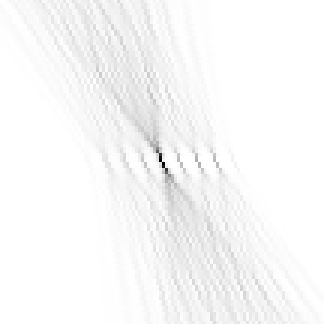}
    \hfill
    \includegraphics[width=0.14\textwidth]{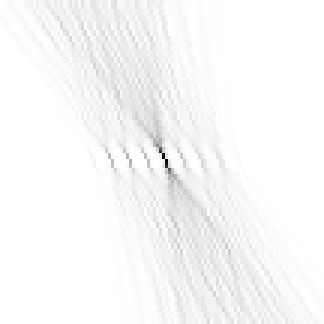}
    \hfill
    \includegraphics[width=0.14\textwidth]{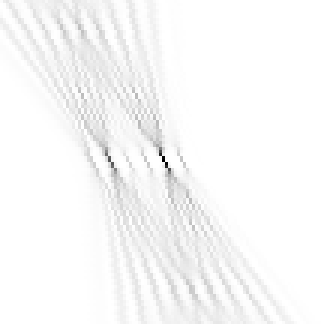}
    \hfill
    \includegraphics[width=0.14\textwidth]{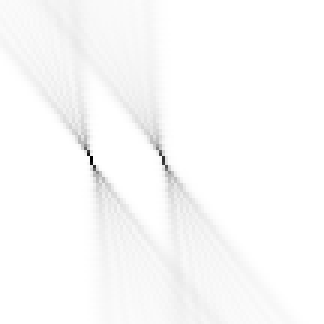}\\
    \makebox[0.8em][c]{\scriptsize $\delta =$}
    \hfill
    {\scriptsize $-89^{\circ}$}
    \hfill
    {\scriptsize $-45^{\circ}$}
    \hfill    
    {\scriptsize $-1^{\circ}$}
    \hfill
    {\scriptsize $+1^{\circ}$}
    \hfill
    {\scriptsize $+45^{\circ}$}
    \hfill
    {\scriptsize $+89^{\circ}$}
    \hfill
    \caption{Example $\fcal(f,\dot{f})$ signal-only patterns in ``grid'' units, for $\iota = 0^{\circ}$ (top row) and $\iota = 180^{\circ}$ (bottom row) and $\delta \in \{ -89^{\circ}, 
    -45^{\circ},
    -1^{\circ}, 
    +1^{\circ}, 
    +45^{\circ}, 
    +89^{\circ}\}$ (from left to right). Centre of each plot corresponds to the injected $f$ and $\dot{f}$, and the maximum of the {\fstat}.}
    \label{fig:FstatDegeneration}
\end{figure}

Figure~\ref{fig:FstatDegeneration} shows example $\fcal$-statistic distributions in the $f$--$\dot{f}$ plane for signals with $\iota = 0$ and $\pi$ (which correspond to maximal values of $h(t)$ and therefore $\fcal$, see Eq.~\ref{eq:ht}), and a range of declination $\delta$ values. Intermediate $\iota$ values result in decreasing $\fcal$-statistic values, with a ``flip'' of $\fcal$-statistic pattern and $\fcal$ reaching its minimum (with the $\times$-component equal 0, see Eq.~\ref{eq:ht}) at $\iota = \pm {\pi}/{2}$. 

\subsection{Data preparation} 
\label{sec:data} 

In the following we briefly describe the input data of the all-sky search pipeline {\tdfstat} \cite{PhysRevD.82.022005,tdfstat-repo}, applied previously to the LVK data \cite{Aasi_2014,PhysRevD.96.062002,PhysRevD.97.102003,PhysRevD.100.024004,PhysRevD.106.102008}. It consists of time series sampled at the sampling time $dt$, duration $T_s$ of several sidereal days long, narrow-banded to a bandwidth $B \equiv 1/(2dt)$. As noted in \cite{1998PhRvD..58f3001J}, a discrete fast Fourier transform (FFT) \cite{Cooley:1965zz} allows to perform the {\fstat} evaluations efficiently. To additionally optimize the coverage in the $f, \dot{f}, \alpha_s, \delta$ parameter space with a minimal number of the matched filter evaluations, the {\tdfstat} implementation  employs a template bank: a grid of pre-selected matched filter parameters. We use a template bank implementation proposed in \cite{2015CQGra..32n5014P,Pisarski_2023}, with a grid of $f, \dot{f}, \alpha_s, \delta$ points, where $f$ values coincide with the FFT frequencies.  We chose the numerical grid (template bank) representation of $\fcal(f,\dot{f})$ as it is convenient for ML training purposes, specifically because, for other parameters fixed, the $\fcal(f,\dot{f})$ distribution pattern in this representation is independent of real (physical) values of $f$ and $\dot{f}$ of the CW signal. Density of the template bank grid in the {\tdfstat} all-sky code, which controls the possible difference in the SNR at adjacent numerical grid points, is a function of the minimal match (MM) parameter, set in our analysis to $\mathrm{MM}=0.99$. 

In this study we focus on CW SIs, i.e.~signals digitally added to the detector's data. To reproduce a realistic scenario, we model our SI CW signals using the parameters of the {\em hardware injections} (HIs), i.e.~signals added on-site by physically actuating detector's hardware \cite{PhysRevD.95.062002,GWOSC_CWHI}. The HIs are important for calibration, validation of instruments and data acquisition systems and, from our point of view, for testing the data analysis methods. We adopt parameters of six CW HIs, commonly denoted as ''pulsars'' (rotating non-axisymmetric NSs), and inject the signals into Gaussian noise to produce software injections (SIs) based on selected ''pulsar'' HIs. Their parameters are summarized in Tab.~\ref{tab:pulsars}.

\begin{table}[t]
    \centering
    \footnotesize 
    \renewcommand{\arraystretch}{1.25}
    \begin{tabular}{|c|c|c|c|c|c|c|c|c|}
    \hline
    Name & 
        \begin{tabular}{@{}c@{}} $h_0$ \\ $[10^{-26}]$ 
        \end{tabular} &
        $\alpha_s$ & $\delta$ &
        \begin{tabular}{@{}c@{}} f \\ $[\mathrm{Hz}]$ \end{tabular} &
        \begin{tabular}{@{}c@{}} $\dot{f}$ \\ $[\mathrm{Hz/s}]$ \end{tabular} &
        $\cos\iota$ &
        %\begin{tabular}{@{}c@{}c@{}} Effective \\ SNR at \\ 6 days \end{tabular} &
        \begin{tabular}{@{}c@{}c@{}} SNR at \\ 6 days \end{tabular} &
        %\begin{tabular}{@{}c@{}c@{}} Effective \\ SNR at \\ 12 days \end{tabular} \\
        \begin{tabular}{@{}c@{}c@{}} SNR at \\ 12 days \end{tabular} \\
    \hline
    P00 & $6.1$ & $4^h46^m$ & $-56.2^{\circ}$ & $265.58$ & $-4.15\cdot10^{-12}$ & $0.795$ & $3.9$ & $5.5$ \\
    P02 & $7.6$ & $14^h21^m$ & $3.4^{\circ}$ & $575.16$ & $-1.37\cdot10^{-13}$ & $-0.929$ & $3.9$ & $5.5$ \\
    P03 & $13$ & $11^h53^m$ & $-33.4^{\circ}$ & $108.86$ & $-1.46\cdot10^{-17}$ & $-0.081$ & $3.9$ & $5.5$ \\
    P05 & $40$ & $20^h10^m$ & $-83.8^{\circ}$ & $52.81$ & $-4.03\cdot10^{-18}$ & $0.463$ & $12.2$ & $17.3$ \\
    P10 & $63$ & $14^h46^m$ & $42.9^{\circ}$ & $26.33$ & $-8.50\cdot10^{-11}$ & $-0.988$ & $12.2$ & $17.3$ \\
    P11 & $32$ & $19^h00^m$ & $-58.3^{\circ}$ & $31.42$ & $-5.07\cdot10^{-13}$ & $-0.329$ & $3.9$ & $5.5$ \\
    \hline
    \end{tabular}
    \caption{Parameters of CW SIs, whose parameters are based on LVK HIs \cite{PhysRevD.95.062002,GWOSC_CWHI}. The amplitudes $h_0$ are originally defined to yield SNR=5 in $100\%$ duty factor 10-day data segments for P00, P02, P03 and P11 injections, and the same value of SNR in 1-day data segments for P05 and P10 injections (see text for details). Here the SNR values are reported for 6 and 12-day long segments.}
    \label{tab:pulsars}
\end{table}

In the following, we consider time domain data in segments of two durations: 6 and 12 sidereal days. The SIs are added into time domain data and then analysed in frequency domain by {\tdfstat} all-sky code. We evaluate the data in frequency bands of width $B=0.5\,\mathrm{Hz}$, which contain frequencies of SIs under study. At a given frequency, detector's noise in time domain is approximated by a standard Gaussian noise with mean $\mu=0$ and the amplitude spectral density (ASD; square root of the PSD $S_f$) equal to $\sigma/\sqrt{\mathrm{Hz}}$, where $\sigma^2$ is the noise variance. In real detector's data, the ASD varies as a function of frequency: for example, in O3 LIGO detectors' data at frequency of $20\,\mathrm{Hz}$ the ASD is approximately $10^{-21}/\sqrt{\mathrm{Hz}}$. Variance of narrow-banded time domain data at frequency $f$ is $\sigma^2 \approx S_fB$, assuming the noise does not evolve as a function of frequency within the band. To obtain a realistic approximation of the real data, we rescale the noise and signal amplitude $h_0$ input data to match the expected SNR of HIs.

Real detector data contain gaps in the time series due to a non-perfect duty factor, bad quality or missing data. It impacts the analysis by reduction of signal's SNR value and suboptimal performance of the matched filter in recovering the $\fcal$-statistic distribution. A dedicated study of this effect is beyond the scope of the current work, although we note here its importance. Nevertheless, we consider here the perfect scenario, when the data segments are complete without any gaps.

\begin{figure} 
    \centering
    \includegraphics[width=0.40\textwidth]
    {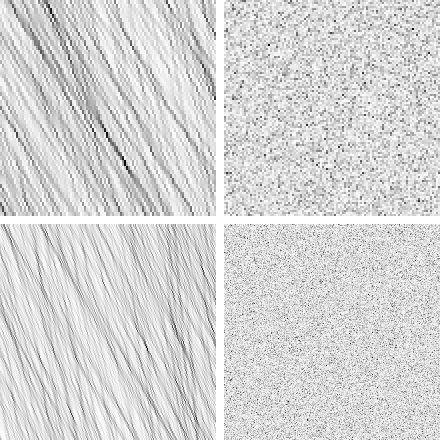}
    \hspace{5pt} 
    \includegraphics[width=0.53\textwidth]{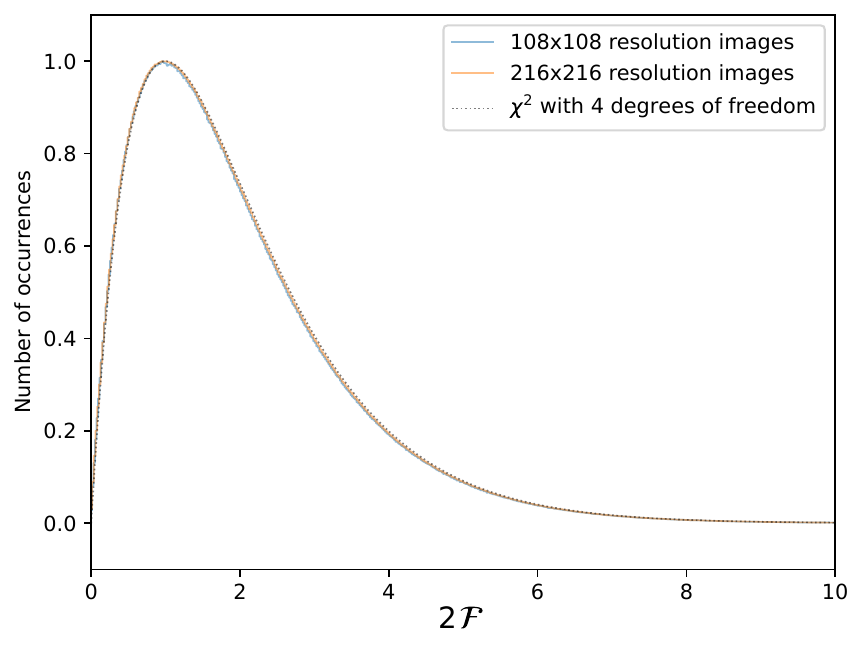}
    \caption{Top row: example of $108{\times}108$ pixels image of $\fcal(f,\dot{f})$, obtained with the {\tdfstat} all-sky pipeline for noise-only 6-days long data segment (left), and a comparison with same-size image obtained by sampling from a $\chi^2_4$ distribution (right). Images are scaled to match the $216{\times}216$ pixel size. Bottom row: same comparison for $216{\times}216$ pixel images (12-day long noise-only data segment). Right panel: distributions of values of pixels for $108{\times}108$ and $216{\times}216$ images for 4000 noise instances, compared with the $\chi^2_4$ distribution, as a function of $2\fcal$. Values are normalised to $[0,1]$ range. Note that despite correlations in $f$--$\dot{f}$ plane expected in the matched-filter output, the distributions are following the standard $\chi^2_4$ distribution.}
    \label{fig:noise_example}
\end{figure}

For the {\tdfstat} search, the length of a given narrow-banded time domain data determines the output resolution in frequency because increasing the $T_s$ increases the resolution of the FFT, and effectively the number of pixels per frequency unit in the $f$--$\dot{f}$ plane. To obtain a single input for our ML implementation, we generate an instance of Gaussian noise, add a generated CW signal for a given sky position, and run the {\tdfstat} search. As a result, a distribution of $\fcal(f,\dot{f})$ values is obtained. We process this information to create images. They are centred at the position of the maximum $\fcal$-statistic value for the signal-only case (the ``ground truth'' position of the {\fstat} pattern; noisy versions of the $\fcal$ distributions are altered by the noise and therefore unreliable for this purpose). The images are trimmed to $108{\times}108$ pixels in case of 6 days segments, and to $216{\times}216$ pixels in case of 12 days segments. Finally, the data is saved as one-channel image normalised in the range $[0, 1]$ to the maximum of the $\fcal$-statistic value in this data instance. Figure~\ref{fig:noise_example} shows exemplary $108{\times}108$ and $216{\times}216$ pixels images of $\fcal(f,\dot{f})$ distributions obtained from the input time domain Gaussian noise for the 6-days segments. Note that these distributions contain non-trivial correlations in the $f$--$\dot{f}$ plane, which are results of the {\fstat} matched filter acting on the Gaussian noise. The overall distribution of $2\fcal$ values is however perfectly compatible with a standard $\chi^2_4$ distribution, as expected \cite{1998PhRvD..58f3001J}. 

\section{Noise removal with machine learning}
\label{sec:MLmethod} 

The post-processing of {\fstat} distributions constitutes a step aimed at recovering the underlying astrophysical signal structure from {\tdfstat} search output corrupted by noise fluctuations. In addition, such a framework may also be used to characterise how consistent the recovered signal is with respect to assumed signal model i.e.~gather information on the significance of the recovery. In the following we describe a ML denoising method stemming from a diffusion-based generative model paradigm \cite{2015arXiv150303585S}. 

\subsection{Machine learning model description}
\label{sec:IADB}

Diffusion models are a class of generative models that learn to synthesise data by modelling the reversal of a gradual noise-injection process \cite{2015arXiv150303585S,2020arXiv200611239H,chen2024overview,GarberTirer,YaoLiZhichangYuming}. Motivated by an analogy with non-equilibrium thermodynamics \cite{2015arXiv150303585S}, the forward process drives a clean data sample toward a noise distribution through a chain of small perturbations, analogous to the physical relaxation of an ordered state toward equilibrium. A model is then trained to reverse the process i.e.~to estimate and remove the noise component from a corrupted sample.

There are several categories of diffusion models, see e.g.~\cite{yang2025,cao2023}. The most common type usually used in image generation are Denoising Diffusion Probabilistic Models \cite{2020arXiv200611239H}, which use a fixed noise schedule to train a NN to reverse the diffusion process. The others are Score-Based Generative Models \cite{10.5555/3454287.3455354,10.5555/3666122.3667767}, which estimate the score function, defined as the gradient of the log probability density of the data; stochastic differential equations are then employed to model the reverse process. Conditional Diffusion Models extend this framework by incorporating additional conditioning information to guide the generation process, with or without classifier guidance \cite{10.5555/3540261.3540933,ho2022classifierfreediffusionguidance}. Discrete Diffusion Models \cite{10.5555/3540261.3541637,10.5555/3540261.3541214} extend the diffusion framework to discrete data, using masking rather than Gaussian noise. Latent Diffusion Models \cite{9878449} perform the diffusion process in a compressed latent space instead the original input space. Cold Diffusion Models \cite{10.5555/3666122.3667911} are further generalizations demonstrating that diffusion-like models can operate with arbitrary corruption processes beyond Gaussian noise. Consistency Models \cite{10.5555/3618408.3619743} ``distill'' diffusion models into generators while preserving generation quality. Finally, Flow Matching \cite{lipman2023flowmatchinggenerativemodeling} generalizes diffusion by learning arbitrary probability paths rather than fixed noise schedules. 

Most of the existing implementations rely on the analytical tractability of Gaussian distribution: a sum of Gaussian random variables is itself Gaussian, while a key requirement in our problem is to produce a model that can effectively treat the $\chi^2$ noise, as well as deal with non-trivial correlations between image points, evident in Fig.~\ref{fig:noise_example}. In addition, the method should be computationally straightforward and stable. Therefore, instead of using {\em probabilistic} diffusion for denoising purposes, we have implemented a {\em deterministic} diffusion: iterative $\alpha$-(de)blending (IADB) method \cite{alpha-deblending}. In contrast to probabilistic approaches which are stochastic in nature \cite{2020arXiv200611239H,10.5555/3454287.3455354}, the IADB constructs a deterministic generative trajectory by linearly interpolating between samples from a source and target distribution, while training a NN to predict the local tangent of the resulting path. 

Suppose $p_0$ and $p_1$ are finite-variance distributions, while $x_0$ and $x_1$ are respective samples from those distributions. We call $\alpha$-blending the linear interpolation mapping between samples $x_0$ and $x_1$, written as 
\begin{equation}
    x_{\alpha} = (1 - \alpha) \, x_0 + \alpha \, x_1 \, ,
\end{equation}
where $\alpha\in[0,1]$. This way we generate samples $x_{\alpha}$ belonging to a distribution $p_{\alpha}$, which is a distribution of samples created by blending samples from $p_0$ and $p_1$ with given $\alpha$; note that given~$x_{\alpha}$ may be generated by different sets of~$(x_0, x_1)$.

The inverse operation -- $\alpha$-deblending -- is a generation of a random pair~$x_0$ and~$x_1$ that produces $x_{\alpha}$. For a given $x_{\alpha}$, the process is carried only over subsets of $p_0$ and $p_1$ distributions. However, for a random~$x_{\alpha}$ all those subsets merge and $\alpha$-deblending occurs over entire $p_0$ and $p_1$ distributions.

We may now construct a deterministic mapping between distributions $p_0$ and $p_1$. Suppose we have a sample $x_{\alpha_1}$ from $p_{\alpha_1}$, deblend it to obtain pair $(x_0, x_1)$, which is a random pair out of all viable pairs which can produce $x_{\alpha_1}$. Then, using parameter $\alpha_2$, we blend this pair to obtain sample $x_{\alpha_2}$ from distribution $p_{\alpha_2}$. If we chain such mappings, gradually increasing $\alpha$ from $0$ to $1$ we obtain a stochastic path connecting $x_0$ from $p_0$ with $x_1$ from $p_1$. Since the path is stochastic then every time we perform such a chain mapping the sample $x_0$ is connected to a different sample $x_1$. However, if we increase the number of steps in our chain mapping then the randomness averages out and we obtain a deterministic mapping between sample $x_0$ and $x_1$, and therefore a deterministic mapping between distribution~$p_0$ and~$p_1$.

We check now if one can navigate between $x_0$ and $x_1$ without explicitly using deblending operation supposing an infinitesimal increase of $\alpha$, denoted by $\Delta\alpha$, which maps $x_{\alpha}$ from $p_{\alpha}$ to $x_{\alpha+\Delta\alpha}$ from $p_{\alpha+\Delta\alpha}$, i.e.
\begin{equation}
    x_{\alpha} \rightarrow (x_0, x_1) \rightarrow x_{\alpha+\Delta\alpha} \, .
\end{equation}
This operation requires choosing one $(x_0, x_1)$ pair which could be blended to obtain sample $x_{\alpha+\Delta\alpha}$ and then infinitesimal increase $\Delta\alpha$ along segment $x_1 - x_0$. However, possible steps average out in the infinitesimally-small increase limit and thus, instead of moving along one of possible $x_1 - x_0$ segments we move along average of those possible segments $\bar{x}_1 - \bar{x}_0$, i.e.
\begin{equation}
    x_{\alpha} \rightarrow (\bar{x}_0, \bar{x}_1) \rightarrow x_{\alpha+\Delta\alpha} \, .
\end{equation}
We can describe small step along chain mapping, $\Delta x_{\alpha}$, using segment obtained from random samples, $x_1 - x_0$, by $\Delta x_{\alpha} = (x_1 - x_0) \, \Delta\alpha$. In the infinitesimal limit where random segments average out we obtain infinitesimal step
\begin{equation}
    \mathrm{d}x_{\alpha} = (\bar{x}_1 - \bar{x}_0) \, \mathrm{d}\alpha \, .
\end{equation}
To predict $\bar{x}_1 - \bar{x}_0$ one can train a NN, $D_{\theta}$. For step $t\in[0, T]$ ($T=1000$ being a sufficiently large number to obtain sufficiently small step, commonly adopted in similar works) we have
\begin{equation}
    D_{\theta}(x_{\alpha_t},\alpha_t) = \bar{x}_1 - \bar{x}_0 \, ,
\end{equation}
while the next sample $x_{\alpha_{t+1}}$ we calculate from the previous one $x_{\alpha_t}$ by
\begin{equation}
    x_{\alpha_{t+1}} = x_{\alpha_t} + (\alpha_{t+1} - \alpha_t) (\bar{x}_1 - \bar{x}_0) \, .
\end{equation}
Our NN learning objective is
\begin{equation}
    \underset{\theta}{\min} \underset{\alpha, x_0, x_1}{\mathbb{E}} \left[ \| D_{\theta} \left( \, (1-\alpha)x_0 + \alpha x_1 \, , \, \alpha \, \right) \, - \, ( x_1 - x_0 ) \|^2 \right] \, .
\end{equation}
In summary, the training and sampling algorithms are presented in pseudo-code as Algs. \ref{alg:training} and \ref{alg:sampling} 
(see Sec.~\ref{sec:results} and \ref{sec:NN_training} for the weighted loss variant used in the actual training). 

\begin{minipage}[t]{0.46\textwidth}
\begin{algorithm}[H]
\small 
\caption{Training}
\begin{algorithmic}
    %\Require $x_0 \in p_0$, $x_1 \in p_1$, $\alpha \in [0,1]$
    %\State $x_{\alpha} = (1-\alpha) \, x_0 + \alpha \, x_1$
    %\State $l = \| D_{\theta}(x_{\alpha},\alpha_ - (x_1 - x_0) \|^2$
    %\State backpropagate from $l$ and update $\theta$
    \Require $x_0 \in p_0$, $x_1 \in p_1$, $T$, $\alpha_t := \frac{t}{T}$
    \For{$t=0,\ldots ,T-1$}
        \State $x_{\alpha_t} = (1-\alpha_t) \, x_0 + \alpha_t \, x_1$
        \State $l = \| D_{\theta}(x_{\alpha_t},\alpha_t) - (x_1 - x_0) \|^2$
        \State backpropagate from $l$ and update $\theta$
    \EndFor
\end{algorithmic}
\label{alg:training} 
\end{algorithm}
\end{minipage}
\hfill
\begin{minipage}[t]{0.46\textwidth}
\begin{algorithm}[H]
\small 
\caption{Sampling}
\begin{algorithmic}
    \Require $x_0 \in p_0$, $T$, $\alpha_t := \frac{t}{T}$
    \For{$t=0,\ldots ,T-1$}
        \State $ x_{\alpha_{t+1}} = x_{\alpha_t} + (\alpha_{t+1} - \alpha_t) \, D_{\theta}(x_{\alpha_t},\alpha_t) $
    \EndFor
\end{algorithmic}
\label{alg:sampling}
\end{algorithm}
\end{minipage} 

\subsection{Neural network implementation}
\label{sec:nn} 

\begin{figure}
    \centering
    \includegraphics[width=1.0\linewidth]{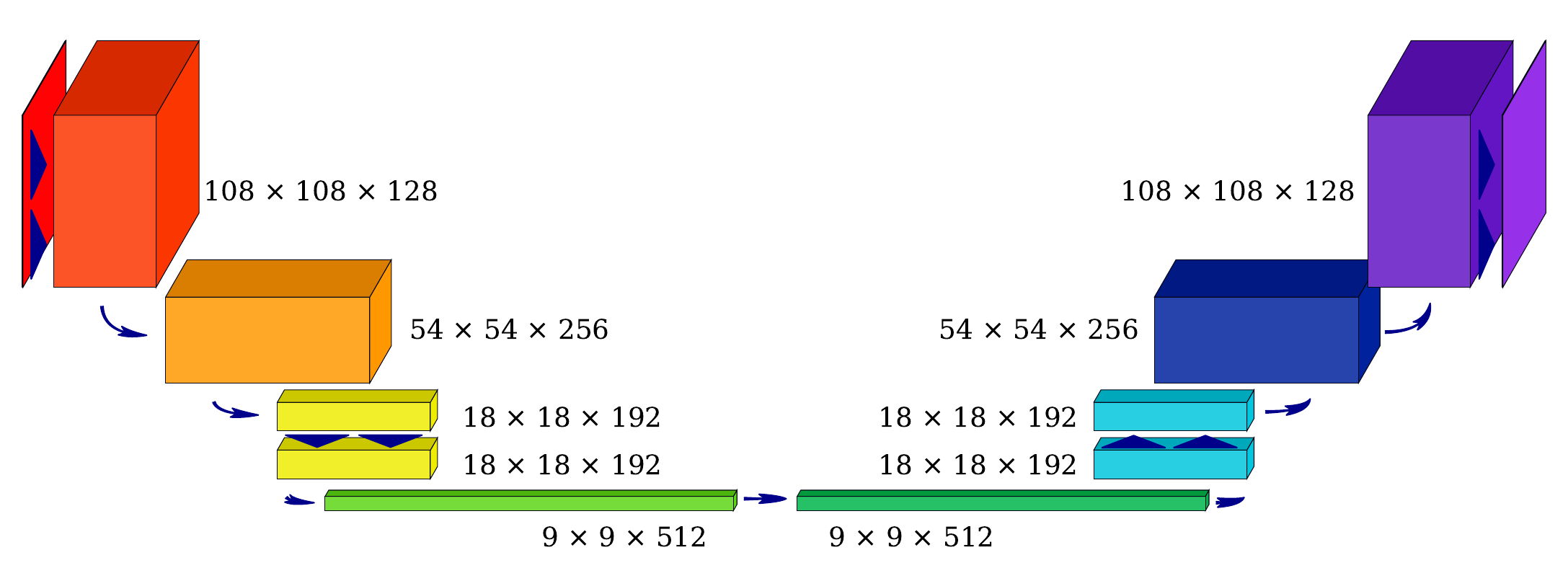}
    \caption{Schematic of the U-Net NN blocks architecture. See text for details.}
    \label{fig:unet}
\end{figure}

We use one main NN architecture to analyse images with two resolutions, $108{\times}108$ and $216{\times}216$ pixels. The images are initially pre-processed with the {\tt maxpool} operation to smooth out but keep the peak values: $108{\times}108$ images with a $2{\times}2$ pixels stride 1 kernel, while for $216{\times}216$ images with $3{\times}3$ pixels stride 2 kernel, to obtain $108{\times}108$ pixels input images. In order to distinguish two models and their results, we call the resulting implementations Model A and Model B, respectively. 

The core NN has the U-Net structure, shown in Fig.~\ref{fig:unet}, with following four convolution down- and up-sampling layers: $108{\times}108$, 128 deep; $54{\times}54$, 256 deep; $18{\times}18$, 192 deep; (attention layer: $18{\times}18$, 192 deep); $9{\times}9$, 512 deep. Spatial resolution gradation is directly motivated by the {\fstat} pattern structure. Main peaks of $\fcal$-statistic signal patterns are distanced in frequency by one sidereal day, i.e.~by~$1.16\cdot10^{-5}\,\mathrm{Hz}$, and resulting separation of the {\fstat} peaks, 6~pixels for 6-day segments and 12~pixels for 12-day segments (which after {\tt maxpool} operation becomes a 6~pixels distance); for details on the training procedure see~\ref{sec:NN_training}.

Network inference process is a procedure of using trained network to perform some specific jobs. In contrast to a general diffusion task of generating data from noise, our models are presented images containing noisy signals, and are expected to remove the noise leaving only the signals' signature, if the signal is present in the data, otherwise produce a blank image. We start the deblending process at the point on chain mapping corresponding to parameter $\alpha$ properly weighting signal and noise images. The higher the SNR of a signal is, the larger parameter $\alpha$.

To properly estimate starting $\alpha$ and initial position in chain mapping we do as follows. For a blended image $\mathbf{P}$ consisting of noise image $\mathbf{N}$ and signal image $\mathbf{S}$ we have 
\begin{equation}
    ( 1 - \alpha ) n + \alpha s = p \, ,
\end{equation}
where, for each pixel position, $n$, $s$, and $p$ are pixel values from $\mathbf{N}$, $\mathbf{S}$ and $\mathbf{P}$, respectively. We can express parameter $\alpha$ by 
\begin{equation}
    \alpha = 1 - \frac{n_{\mathrm{max}}}{p_{\mathrm{signal}}} \, .
\end{equation}
where $n_{\mathrm{max}}$ is the maximal pixel value (in {\fstat} values) from $\mathbf{N}$, and $p_{\mathrm{signal}}$ is the pixel value from $\mathbf{P}$ at the top peak pixel position of the {\fstat} signal function. Signal images are centred on the maximal {\fstat} value so the value of images centre pixel is always $p_{\mathrm{signal}}$. In principle, noise values do not have an upper limit. However, in practice averaging out many noise instances we find that $n_{\mathrm{max}} \approx b \, \tilde{n}$, where~$\tilde{n}$ is the median noise pixel value, and $b$ is a scaling factor calculated separately, in our case averaged over sufficiently large number of instances $N$ (in our case $N=4000$). The number of instances used here matches the number of denoising attempts for each pulsar during inference (see below). Keeping equal in number sample base for calculation of~$b$ and for inference allows us to
use the calculated~$b$ value for computing parameter~$\alpha$ during the inference. Moreover, addition of signal to the noise is negligible for calculation of median pixel value of $P$ so we have $\tilde{p} \approx \tilde{n}$. Finally we obtain
\begin{equation}
\label{eq:alpha}
    \alpha = 1 - \frac{b \, \tilde{p}}{p_{\mathrm{signal}}} \, .
\end{equation}
In practice, it may happen that $\alpha$ calculated using Eq.~\ref{eq:alpha} is negative. In those cases we simply set its value to 0. The actual values of scaling factor $b$ we calculate analysing $N=4000$ instances of noise after applying the {\tt maxpool} operation. These values are $b_{6\mathrm{days}} = 3.9$ and $b_{12\mathrm{days}} = 3.7$.
%$b_{6\mathrm{days}} = 3.856$ and $b_{12\mathrm{days}} = 3.650$

Real case scenario brings the issue of factor $b$ estimation. A typical observing run duration of the LVC is of the order of a year, resulting in less than a hundred unique multiple-days long segments. Additionally, one can shift the segments beginning and end times to create partially overlapping input. This should result in up to a few hundred,~$N$, images containing different instances of candidate signal within noise which can be deblended using our ML model. Then using a real detector ASD of given frequency band one can generate~$N$ pure noise instances and use them to estimate factor~$b$ using~$n_{\mathrm{max}} \approx b \, \tilde{n}$ relation. And this in turn allows to calculate parameter~$\alpha$ for each candidate signal within noise image.

Assessing the similarity of two images is generally a difficult task. We have experimented with straightforward metrics like the root mean squared error summed over all pixel values, and with the Learned Perceptual Image Patch Similarity (LPIPS) metric implemented in the \texttt{lpips} library \cite{LPIPS}, which benefits from a pre-trained NN (e.g. the AlexNet~\cite{AlexNet}). Eventually for the results presented below, we adopted the Structural Similarity Index Measure (SSIM) metric \cite{SSIM_ref1,SSIM_ref2}, which performs exceptionally well when evaluating structured outputs such as images. It combines analysis of image luminance, contrast, and structures. Moreover, contrary to LPIPS, SSIM is not human-perception biased. In practical terms, as a result of comparing two images the SSIM metric yields a value in a $[-1,1]$ range, which quantify the similarity of the images. Value of~$1$ means perfect match, while~$-1$ denotes perfectly opposite image (i.e.~a negative).

\section{Results}
\label{sec:results}

\begin{figure}
    \centering
    \includegraphics[width=0.49\linewidth]{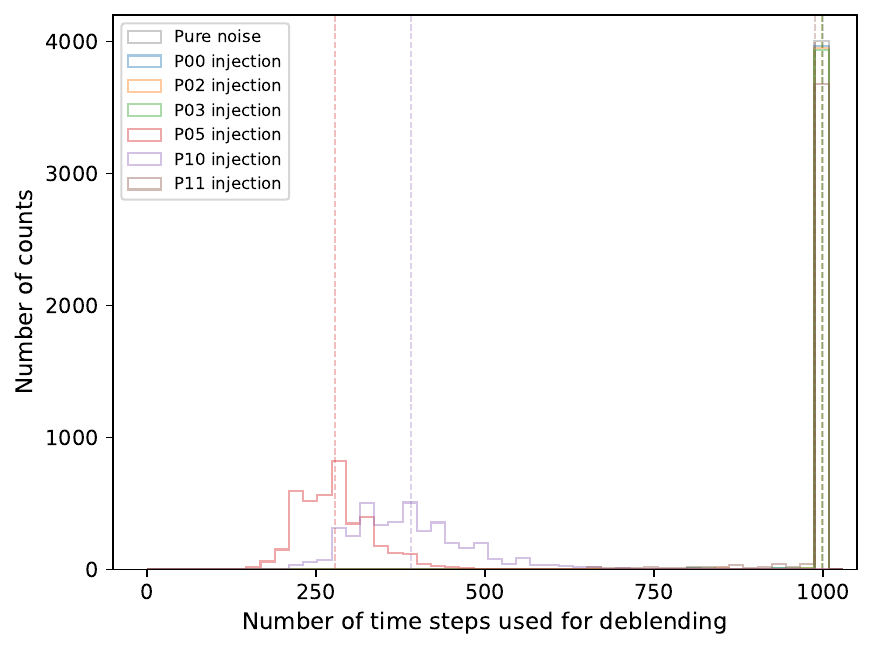}
    \includegraphics[width=0.49\linewidth]{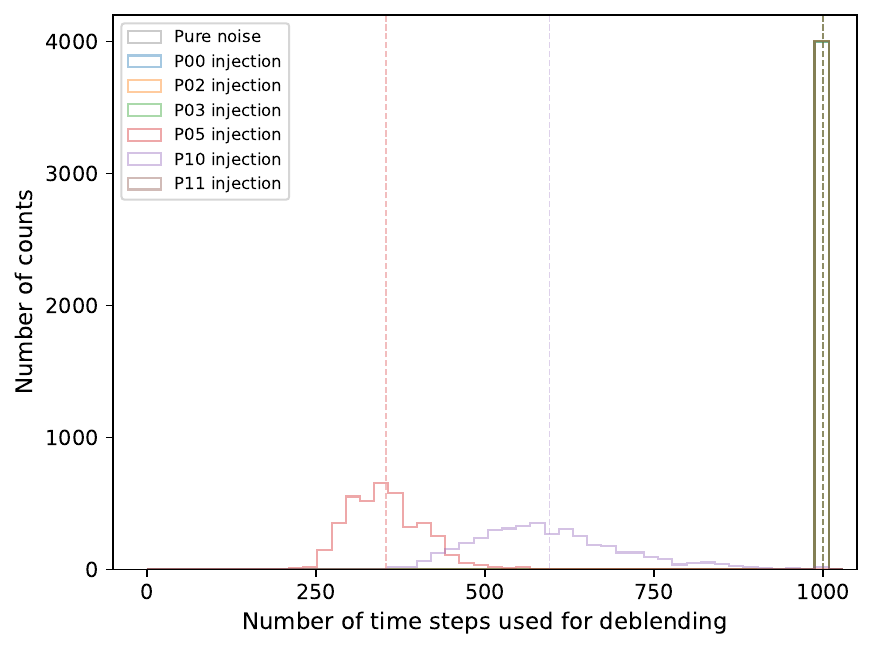}
    \caption{Number of steps required for deblending for Model~A (left) and Model~B (right), guided by Eq.~\ref{eq:alpha}. Vertical dashed lines mark average number of steps values.}
    \label{fig:steps}
\end{figure}

Depending on the SI's amplitude $h_0$, the procedure may require fewer than the maximum of 1000 deblending steps. Using the estimate from previous section, in Fig.~\ref{fig:steps} we show histograms of number of steps required for deblending in 4000 instances for all SIs for both models. As expected, high SNR SIs (P05 and P10) require less deblending steps. 

To verify the robustness of the results and to check that the method has properly recovered the ''ground truth'' {\fstat} pattern, we compare deblended images with template signal images and with a blank image. As shown in Sec.~\ref{sec:fstat_pattern}, Fig.~\ref{fig:FstatDegeneration} and \ref{sec:fstat_degeneracy}, the {\fstat} pattern image is degenerate with respect to simultaneous changes in $\delta$ and $\iota$. We note here that the training and the ''ground truth'' data set assumes $\iota=0$ or $\iota=\pi$, whereas the SIs cover a wide range of this parameter. In the following tests we focus on the $\delta$ dependence, since it results in the most visible differences, e.g.~the number and relative heights of the peaks. In real case scenario of a directed search, the ''ground truth'' $\delta$ is known, so the results presented here may be used as a consistency/veto check for recovered versus expected signal pattern.

The test for robustness of the method consists of generating 4000 images for each SI, based on $N=4000$ randomly generated time domain noise data for each SI, evaluating these input data with the {\tdfstat} pipeline to create $\fcal(f,\dot{f})$ images, and calculating the $\alpha$ parameter related to the optimal number of steps to deblend each image. 
Next, for each SI, each of the $N$ deblended versions is compared with the set of template signals ({\fstat} patterns as functions of signal parameters, see e.g.~Fig.~\ref{fig:FstatDegeneration}), and the SSIM metric is calculated. The largest the output value, the closer the deblended image to the ''ground truth'' image: assessing the largest SSIM value serves as a proxy to estimate the SIs ''ground truth'' parameters, in this case the declination angle $\delta$. We perform the evaluation using images which were deblended to non-blank image only. For each SI, we find the SSIM value averaged across all the $N$ trial instances with all possible ''ground truth'' signal patterns, and thus we obtain SSIM value distribution as a function of the parameter $\delta$.

\begin{figure}
    \centering
    \begin{subfigure}{1.0\textwidth}
    \centering
        \includegraphics[width=1.0\linewidth]{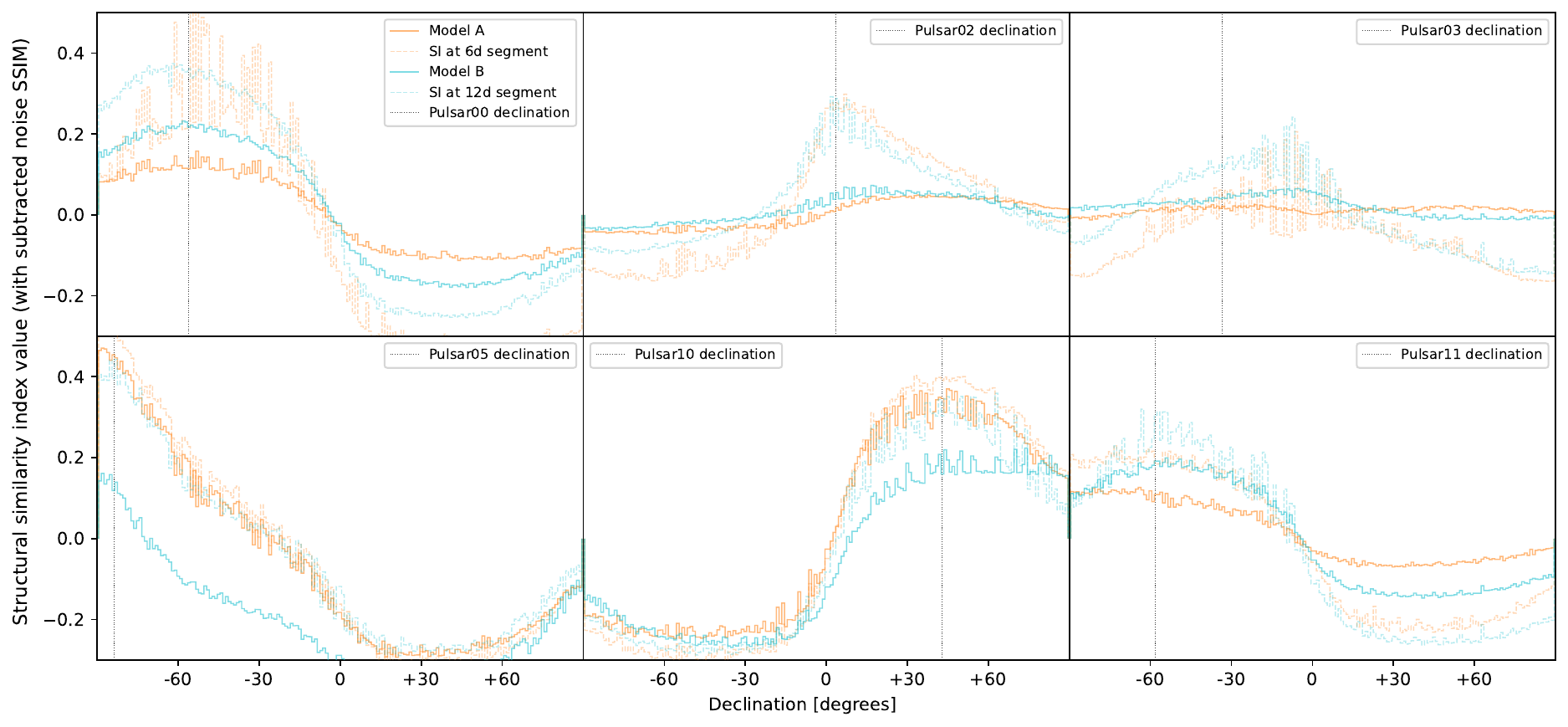}
    \end{subfigure}
    \caption{Distributions of SSIM values for \textit{SIs \& noise} (solid lines) and \textit{pure SIs} (dashed lines) \textbf{with pixel-wise subtracted noise distribution} (Fig.~\ref{fig:ResultsNoise}). Larger values denote higher similarity. Vertical lines mark the ''ground truth'' value of SIs declinations. Table~\ref{tab:FractionsMain} contains results for deblended non-blank fraction of each SI.}
    \label{fig:ResultsModelAandB}
\end{figure}

Furthermore, to take into account the bias due to noise, we calculate analogous SSIM value distribution but for deblended images of just noise data instances, i.e.~without any SI signal added. Then, declination bin by declination bin, we subtract the obtained \textit{noise} SSIM value distribution from the \textit{SI \& noise} SSIM value distributions. The results are shown in Fig.~\ref{fig:ResultsModelAandB}. Dotted vertical lines denote the ''ground truth'' declination $\delta$ for a given SI. Clustering of SSIM value around the highlighted declination indicates that the model has correctly deblended the SI.

In Fig.~\ref{fig:ResultsModelAandB} we also plot the SSIM value distributions obtained by comparing pure SI signals (i.e.~without any noise added) with all template signals. Also from those distributions we subtract \textit{noise} SSIM distribution. Dashed lines represent cases when all \textit{SI \& noise} images would be deblended to respective exact SI signal image. This however can not be realised due to the fact that in our template signal images there is no any exact SI signal image. Thus, our NN tries deblending noisy images to template signal images ''closest'' to the exact SI images. This is the reason why the SSIM value distributions might never achieve exact SI profiles having the distributions fit depending on the difference between template signal and exact SI images.

\begin{figure}
    \centering
    \begin{subfigure}{1.0\textwidth}
    \centering
        \includegraphics[width=0.49\linewidth]{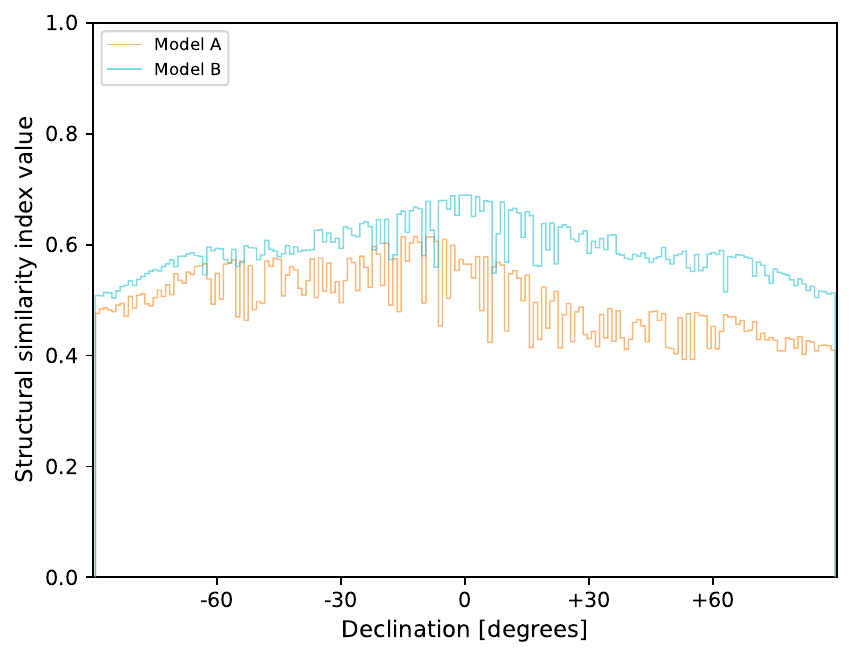}
    \end{subfigure}
    \caption{Deblending results of images containing only noise for both A and B models.}
    \label{fig:ResultsNoise}
\end{figure}

For comparison, Fig.~\ref{fig:ResultsNoise} shows deblending results for images containing \textit{only noise} which produce roughly flat noise distributions of SSIM values as a function of $\delta$. These distributions we use to adjust \textit{SI \& noise} SSIM distributions~(cf.~Fig.~\ref{fig:ResultsModelAandB}). Deblending of \textit{noise only} images results in roughly 20\% of blank output images which is consistent with the amount of blank images in the set of training signal template images. Deblending of \textit{SI \& noise} images always produce more signal images than in case of deblending \textit{noise only} images. This fraction can reach even 100\% for high SNR signals, regardless whether the SSIM distribution properly identifies the signal or not, see Tab~\ref{tab:FractionsMain}.

\begin{table}[t]
    \centering
    \footnotesize 
    \renewcommand{\arraystretch}{1.25}
    \begin{tabular}{|c|c|c|c|c|c|c|c|}
    \hline
    & Noise & P00 & P02 & P03 & P05 & P10 & P11 \\
    \hline
    Model A (6 days) & $0.84$ & $0.92$ & $0.89$ & $0.87$ & $1.00$ & $1.00$ & $0.91$ \\
    Model B (12 days) & $0.71$ & $0.98$ & $0.91$ & $0.91$ & $0.79$ & $1.00$ & $0.93$ \\
    \hline
    \end{tabular}
    \caption{Fractions of images deblended to non-blanks for results in Figs.~\ref{fig:ResultsModelAandB} and \ref{fig:DistributionsSIs}.}
    \label{tab:FractionsMain}
\end{table}

\begin{figure}
    \centering
    \begin{subfigure}{1.0\textwidth}
    \centering
        \includegraphics[width=1.0\linewidth]{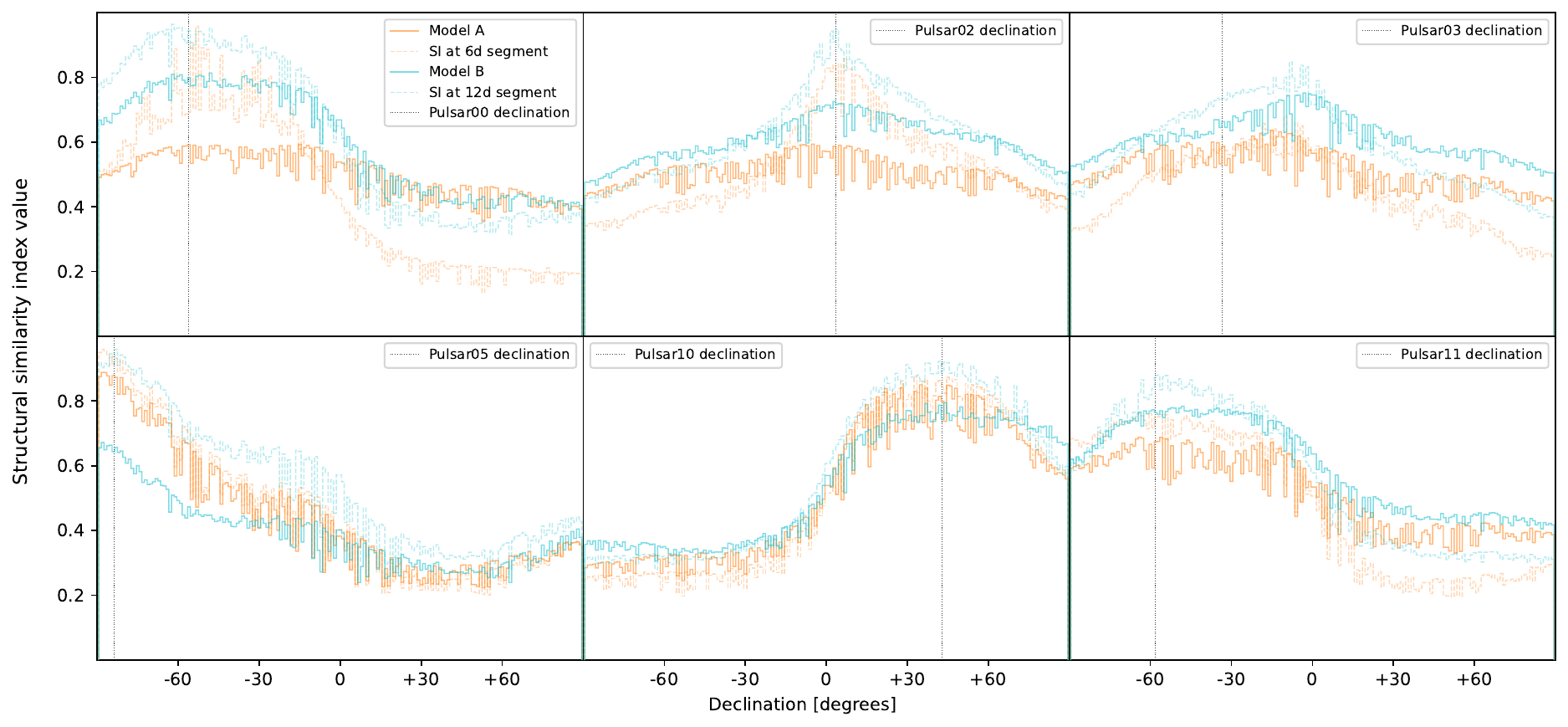}
    \end{subfigure}
    \caption{Distributions of SSIM values for \textit{SIs \& noise} (solid lines) and \textit{pure SIs} (dashed lines). Larger values denote higher similarity. Vertical lines mark the ''ground truth'' value of SIs declinations. Table~\ref{tab:FractionsMain} contains results for deblended non-blank fraction of each SI.} 
    \label{fig:DistributionsSIs}
\end{figure}

In Fig.~\ref{fig:DistributionsSIs} we plot deblended \textit{SI \& noise} images SSIM value distributions as well as \textit{pure SI} SSIM value distributions without any noise distribution adjustment. \textit{Pure SI} distributions clearly show that analysis of longer time segments confer higher precision in comparing images, i.e.~higher SSIM values.

Plots in Fig.~\ref{fig:ResultsModelAandB} corrected for ''background noise'' show significance of network detections. In case of P02 and P03 SIs the networks output SSIM distributions are roughly flat with SSIM values around 0 for both models. In case of remaining SIs the SSIM distributions attain profiles in-line with \textit{pure SI} distributions. It allows us to assert that in those cases our model correctly detects the presence of a signal hidden in noise.

Figure~\ref{fig:DistributionsSIs} shows a direct comparison  results between models and \textit{pure SIs}. Similarly to \textit{pure SIs} distributions we note that in case of both models distributions, longer time segments in most cases provide higher SSIM values as well as a greater contrast between peak and lowest distribution values.

\begin{figure}
    \centering
    \begin{subfigure}{1.0\textwidth}
    \centering
        \includegraphics[width=0.3\linewidth]{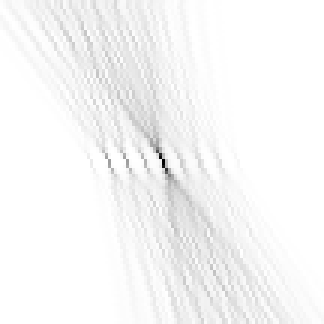}
        \includegraphics[width=0.3\linewidth]{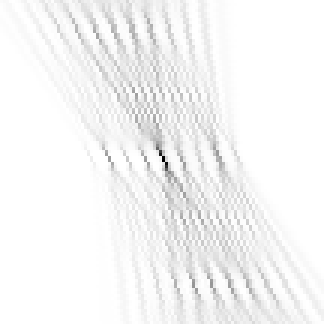}
    \end{subfigure}
    \caption{{\fstat} image of P02 SI for 6d segment data (left panel) and P03 SI for 6d segment data (right panel).}
    \label{fig:P02andP03SI}
\end{figure}

\begin{figure}
    \centering
    \begin{subfigure}[][][c]{0.24\textwidth}
        \includegraphics[width=\linewidth]{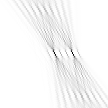}
    \end{subfigure}
    \begin{subfigure}[][][c]{0.37\textwidth}
        \includegraphics[width=\linewidth]{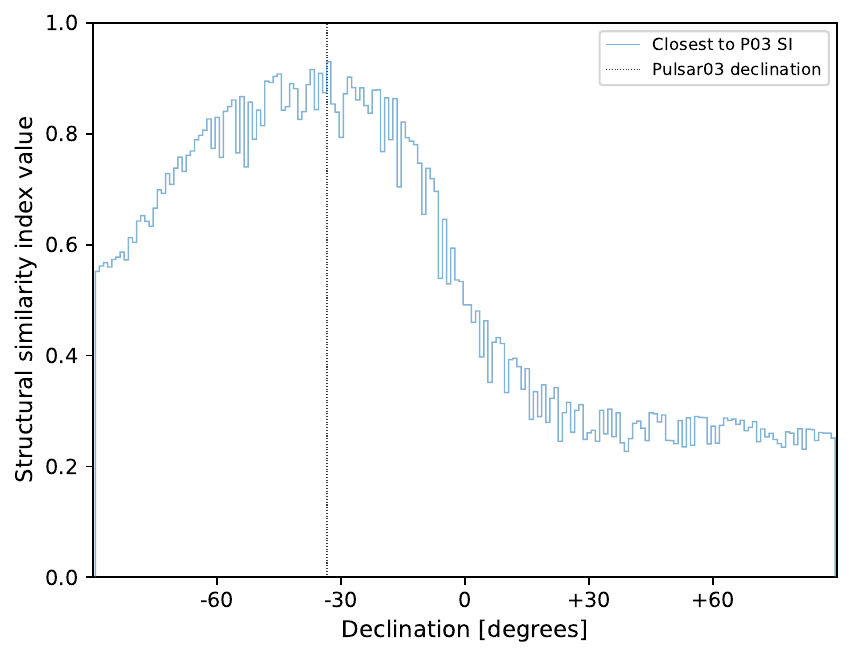}
    \end{subfigure}
    \begin{subfigure}[][][c]{0.37\textwidth}
        \includegraphics[width=\linewidth]{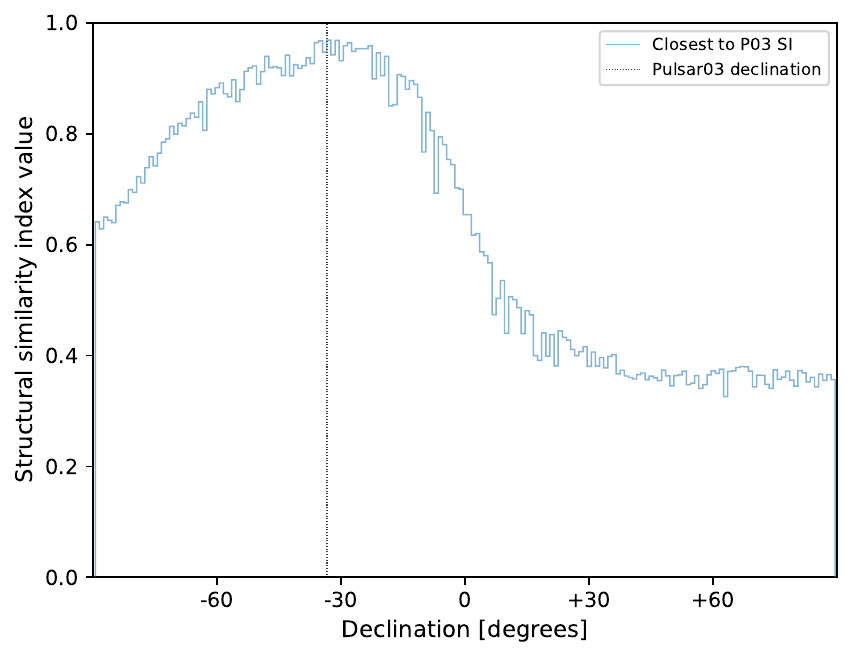}
    \end{subfigure}
    \caption{Signal template image closest to the P03 SI for 6d data segment (cf.~Fig.~\ref{fig:P02andP03SI}), with $\delta = -33^{\circ}$ and $\iota=\pi$ (left panel). SSIM values distributions between template signal image closest to P03 SI and all template signal images for 6d segment data (middle panel) 12d segment data (right panel).}
    \label{fig:ClosestP03}
\end{figure}

We attribute the sub-standard performance in detecting P03 SI to near zero value of this signal's~$\cos\iota$. Normally, this nearly linearly polarised signal should be very weak but is boosted in~$h_0$ value (in relation with respective ASD) for the injection purposes. Its {\fstat} has a low ratio between maximal and minimal {\fstat} peak values and thus in the image pixel value normalisation between $[0,1]$ the image has a form of many comparably saturated peaks, see Fig.~\ref{fig:P02andP03SI}. This makes it qualitatively distinct from other template signal images and is difficult to properly denoise by our network. The training template signal images closest to P03 SI are templates for~$\delta = -33^{\circ}$ and $\iota = \{0,\pi\}$. They essentially match the true injection declination ($\delta=-33.4^\circ$, see Tab.~\ref{tab:pulsars}), so the resulting mismatch is in inclination, with P03 SI $\,\iota\,{\approx}\,\pi/2$, roughly equidistant from both template extremes. Thus P03 SI true morphology is almost perfectly out-of-distribution for a template bank. Plots in Fig.~\ref{fig:ClosestP03} demonstrate the calculated SSIM distributions. 

\begin{figure}
    \centering
        \centering
        \includegraphics[width=0.49\linewidth]{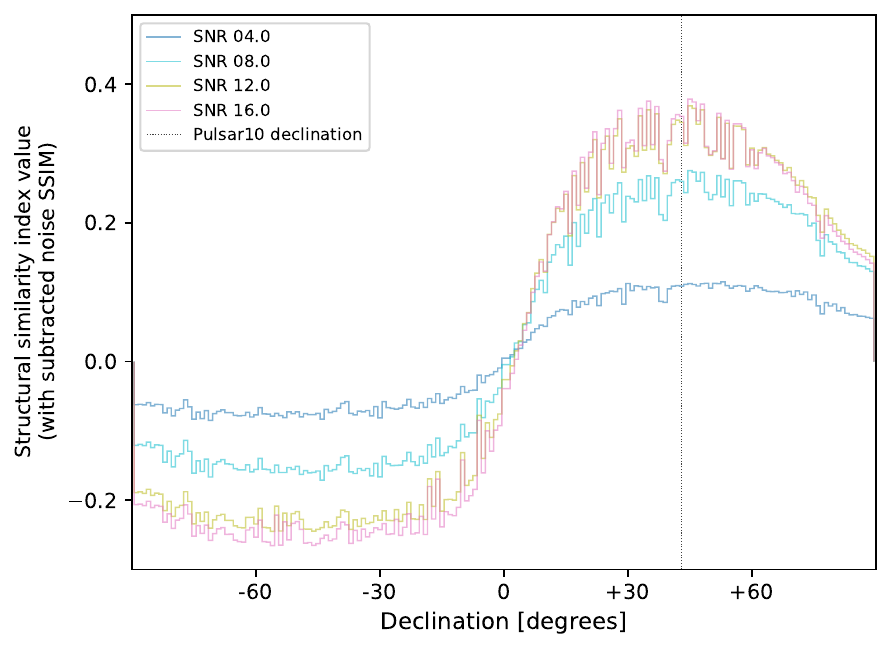}
        \includegraphics[width=0.49\linewidth]{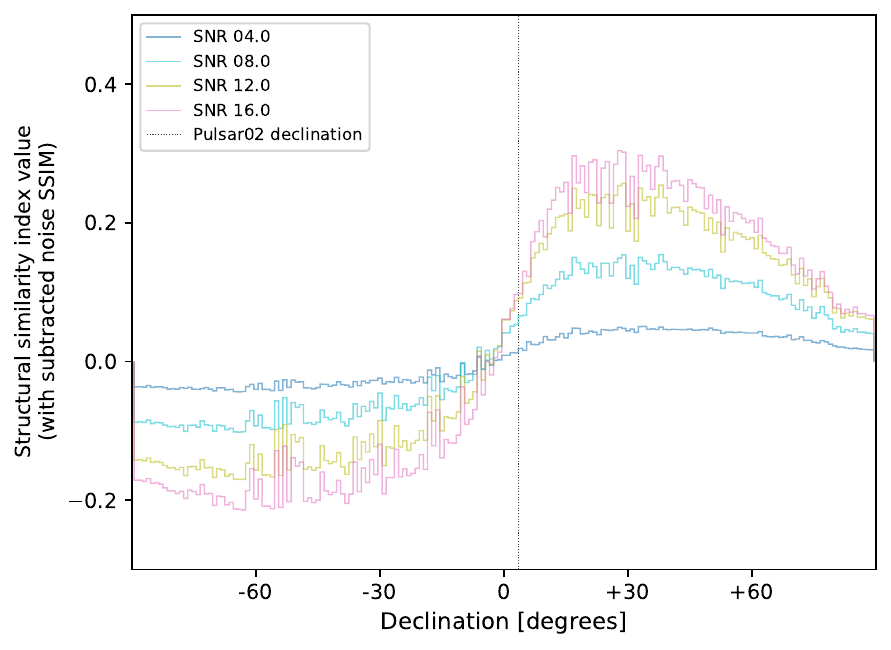}
    \caption{Deblending results images of P10 SI (left panel) and P02 SI (right panel), with varying effective SNR for Model~A.}
    \label{fig:ResultsManySNR}
\end{figure}

The P02 SI result is investigated in a simple study of the SNR effect on deblending. In particular, for a 6 days long time segment we produce a set of P10-like SIs varying in SNR. After deblending 4000 instances of each SI we obtain expected result, i.e.~increase in SNR produces more pronounced clustering of SSIM value distribution near the ''ground truth'' declination, see left panel in Fig.~\ref{fig:ResultsManySNR}. Similar results are obtained with deblended sets of P02 SI images with progressively higher SNRs. However, even though finding a signal, the method fails to recover the correct declination value, see right panel in Fig.~\ref{fig:ResultsManySNR}. We attribute this result to a particularly difficult, from the point of view of deblending, SI image pattern. The declination of the source is $\delta\approx 0^\circ$, which results in a solitary {\fstat} peak, see Fig.~\ref{fig:P02andP03SI}, and may be easily confused with matched-filter noise artifacts, or parts of the {\fstat} patterns at other declinations. Fractions of images deblended to non-blank image in case of P02 and P10 SIs can be found in Tab.~\ref{tab:FractionsManySNR}. Again, this result shows that with the increase of signal SNR the fraction of non-blank images rise above the pure noise case fraction level eventually reaching 100\%.

\begin{table}[t]
    \centering
    \footnotesize 
    \renewcommand{\arraystretch}{1.25}
    \begin{tabular}{|c|c|c|c|c|}
    \hline
    & SNR 4.0 & SNR 8.0 & SNR 12.0 & SNR 16.0 \\
    \hline
    P02 & $0.88$ & $0.96$ & $0.99$ & $1.00$ \\
    P10 & $0.91$ & $0.99$ & $1.00$ & $1.00$ \\
    \hline
    \end{tabular}
    \caption{Fractions of images deblended to non-blanks for results presented in Fig.~\ref{fig:ResultsManySNR}.}
    \label{tab:FractionsManySNR}
\end{table}

\section{Discussion and conclusions} 
\label{sec:summary}

Our approach to generative ML in the context of GW data analysis is a novel application to signal post-processing. The results presented above show that the {\tdfstat} all-sky search output is a suitable target for deterministic generative denoising: the noisy $\fcal(f,\dot{f})$ images contain non-trivial spatial correlations inherited from the matched filter while their pixel-value distributions remain consistent with the expected $\chi^2_4$ distribution, validating the choice of the deterministic IADB approach over standard probabilistic diffusion models, which rely on Gaussian-noise assumptions. The number of deblending steps required to recover a candidate tracks its SNR, with high-SNR injections converging in markedly fewer steps than weak ones. 

Comparison of deblended images against the library of possible {\fstat} template patterns using the SSIM metric shows that the method reliably clusters around the correct declination for $\rho \gtrsim 4$, except in specific cases discussed below. This threshold is estimated with multiple injections with varying SNR, see Fig.~\ref{fig:ResultsManySNR} with the lowest tested value $\rho = 4$. Four of the six hardware-injection-based SIs (P00, P02, P03, P11) have the same SNR $\rho\,{\approx}\,4$, yet only half of them (P00, P11) are correctly recovered, indicating that at SNR near this threshold, recovery is governed primarily by signal morphology. Weaker signals, and pure noise instances, produce uninformative, flat SSIM distributions, indicating the method does not manufacture spurious structures. In those cases the fraction of images deblended to signal image is consistent with training template signal fraction, while it grows with increasing signal SNR. 

The recovery is not governed by the SNR alone. The P02 SI, whose declination $\delta\approx0^\circ$ produces a single {\fstat} peak, is found, yet does not reliably unravel the correct declination value, since this morphology is easily confused with the matched-filter noise artifacts. Furthermore, we interpret the sub-standard deblending results of the P03 SI by its $\cos\iota$ value which makes its {\fstat} image in our processing method quite distinct from the template {\fstat} images. This ``corner-case result'' is important in marking the boundary of the extrapolation capabilities of the method.  

Comparing Model~A (6-day segments) and Model~B (12-day segments) further shows that longer coherence time does not uniformly improve recovery: Model~A recovers P05 SI more completely than Model B (non-blank fraction 1.00 vs. 0.79, see Tab.~\ref{tab:FractionsMain}). This is notable given P05 SI high SNR, and suggests that factors beyond duration and amplitude (possibly small spindown magnitude and/or its declination near the pole) also influence recoverability.

These findings motivate several extensions of the present study. A natural one is to move from single-segment denoising toward sequential analysis across consecutive (and potentially overlapping) segments of a full observing run, integrating the method more closely with the semi-coherent coincidence stage of a real search. One should also investigate the idealised assumptions adopted here, namely the absence of data gaps and the flat noise power spectral density within the analysis band, in favour of realistic detector duty factors and frequency-dependent noise. 

The failure of correct $\delta$ identification for the single-peak mode or low-declination sources warrants dedicated investigation, whether through richer training coverage of parameter space, architectural changes to the network, or additional conditioning information that could help disambiguate genuine weak signals from noise artifacts.

Due to the nature of our method, in particular the inspection of {\fstat} images at a specific frequency-spindown point, the method is the best suited to directed searches, or an outlier follow-up stage after the first stage of an all-sky search, as the computational cost of evaluating every point in frequency-spindown space is prohibitive.

Further work could also broaden the training process and testing parameter space beyond the six hardware-injection-based pulsars considered here, explore alternative or complementary similarity metrics beyond SSIM, possibly combined with the deblending step count as a joint significance proxy, and consider ensemble or probabilistic variants of IADB to obtain uncertainty estimates rather than single deterministic outputs. Finally, integrating the denoising/veto step directly into the {\tdfstat} search pipeline is a potential follow-up improvement.

{\ack
This work was supported by the Polish National Science Centre grants No. 2023/49/B/ST9/02777 and 2021/43/B/ST9/01714, and by the European Union-Next Generation EU, Mission 4 Component 1 CUP J53D23001550006 with the PRIN Project No. 202275HT58. Computing resources were partially provided by the Nicolaus Copernicus Astronomical Center of the Polish Academy of Sciences.
}

% Appendices 
\appendix

\section{\texorpdfstring{$\fcal$}{F}-statistic dependency on angle parameters}
\label{sec:fstat_degeneracy}

We start with noting that the signal's amplitude $h(t)$ depends on the source's sky position through the antenna patterns (Eq.~\ref{eq:ht}). For the data duration $T_s$ of  multiple sidereal days, the antenna pattern dependence on right ascension $\alpha_s$ averages out due to Earth's rotation, making the $\fcal$ value independent of $\alpha_s$. Effectively, signals with a given $\delta$ but different $\alpha_s$ are phase-shifted versions of one another; for $T_s$ of multiple sidereal days, meaning full $2\pi$ revolutions, this phase shift is irrelevant; see Appendix B of \cite{PhysRevD.65.042003}. In the following we examine how $h(t)$ changes under simultaneous transformation $\delta \to -\delta$ and $\iota \to \iota+\pi$. We rewrite Eq.~\ref{TotalH} as in \cite{1998PhRvD..58f3001J} 
\begin{equation}
\label{AppendixEqH}
h(t) = F_+ (t;\delta,\psi,\ldots) \, h_+(t;\iota,\phi) + F_\times (t;\delta,\psi,\ldots) \, h_\times(t;\iota,\phi) \, ,
\end{equation}
where $h_+$ and $h_\times$ are two independent wave polarization functions. Those functions are independent of $\delta$, 
\begin{equation}
\label{AppendixEqHplusminus}
\begin{aligned}
h_+ (t;\iota+\pi,\phi) &= \frac{1+\cos^2(\iota+\pi)}{2}\cos\phi(t)
= \frac{1+\cos^2\iota}{2}\cos\phi(t) = h_+ (t;\iota,\phi) \, ,\\
h_\times (t;\iota+\pi,\phi) &= \cos(\iota+\pi)\sin\phi(t) = -\cos\iota\sin\phi(t)
= -\,h_\times (t;\iota,\phi) \, ,
\end{aligned}
\end{equation}
whereas the detector antenna pattern are 
\begin{equation}
\begin{aligned}
F_{+}(t;\delta,\psi,\ldots)
  &= a(t;\delta,\ldots)\,\cos 2\psi
   + b(t;\delta,\ldots)\,\sin 2\psi \, , \\
F_{\times}(t;\delta,\psi,\ldots)
  &= b(t;\delta,\ldots)\,\cos 2\psi
   - a(t;\delta,\ldots)\,\sin 2\psi \, ,
\end{aligned}
\end{equation}
with explicit formulas for~$a$ and~$b$ functions as in Eqs.~(12) and~(13) in \cite{1998PhRvD..58f3001J}. Considering dependence of~$a$ and~$b$ on $\delta$ we note that
\begin{equation}
\begin{aligned}
a(t; \delta, \ldots)
  &\sim C_1 \cos\delta
      + C_2 \cos\delta\,\cos 2\delta
      + C_3 \cos^3\delta
      + \ldots \, , \\
b(t; \delta, \ldots)
  &\sim D_1 \sin\delta
      + D_2 \sin\delta\,\cos^2\delta
      + D_3 \sin 3\delta
      + \ldots \, ,
\end{aligned}
\end{equation}
where $C_i$ and $D_i$ functions depend on time, but not $\delta$. Moreover, $a$ is built from even functions of $\delta$, while dependence of $b$ on $\delta$ always carries an odd function equivalent. Therefore
\begin{equation}
a(t; -\delta, \ldots)=a(t; \delta, \ldots)\quad\text{and}\quad 
b(t; -\delta, \ldots)=-\,b(t; \delta, \ldots) \, .
\end{equation}
Note that $\psi$ is defined relative to a sky basis that itself changes under $\delta \to -\delta$ transformation and thus that $\delta$ transformation \textit{forces} sign change in $\psi$ as well. We can derive now the impact of the flip (sign change) in $\delta$ on antenna patterns
\begin{equation}
\label{AppendixEqFplus}
\begin{aligned}
F_{+}(t;-\delta,-\psi,\ldots) &= a(t;-\delta,\ldots) \, \cos(-2\psi) + b(t;-\delta,\ldots) \, \sin(-2\psi) \\
    &= a(t;\delta,\ldots) \, \cos(2\psi) + b(t;\delta,\ldots) \, \sin(2\psi) \\
    &= F_{+}(t;\delta,\psi,\ldots)
\end{aligned}
\end{equation}
\begin{equation}
\label{AppendixEqFminus}
\begin{aligned}
F_{\times}(t;-\delta,-\psi,\ldots) &= b(t;-\delta,\ldots) \, \cos(-2\psi) - a(t;-\delta,\ldots) \, \sin(-2\psi) \\
    &= -b(t;\delta,\ldots) \, \cos(2\psi) + a(t;\delta,\ldots) \, \sin(2\psi) \\
    &= -F_{\times}(t;\delta,\psi,\ldots)
\end{aligned}
\end{equation}
Finally, with Eqs. \ref{AppendixEqHplusminus}, \ref{AppendixEqFplus} and \ref{AppendixEqFminus}, $h(t)$ is
\begin{equation}
\begin{aligned}
&h(t;\iota+\pi,-\delta,-\psi,\ldots) = \\ 
&=F_+ (t;-\delta,-\psi,\ldots) \, h_+(t;\iota+\pi,\phi) + F_\times (t;-\delta,-\psi,\ldots) \, h_\times(t;\iota+\pi,\phi) \\ 
&=F_+ (t;\delta,\psi,\ldots) \, h_+(t;\iota,\phi) - F_\times (t;\delta,\psi,\ldots) \cdot \left(-h_\times(t;\iota,\phi)\right) \\
&=h(t;\iota,\delta,\psi,\ldots) \, .
\end{aligned}
\end{equation}
Hence, under simultaneous transformation $\iota \to \iota+\pi$ and $\delta \to -\delta$ (and \textit{forced} $\psi \to -\psi$) the signal function $h(t)$ is invariant and, subsequently, also the $\fcal$-statistic of this signal, 
\begin{equation}
\fcal(\delta, \iota, \psi) = \fcal(-\delta, \iota+\pi, -\psi) \, .
\end{equation}

\section{Neural network training procedure}
\label{sec:NN_training}

The training datasets contain 52388 noise images for Model~A and 48956 noise images for Model~B. Template signal image sets contain the same number of images in both model cases, i.e.~356 signal images for various parameters and 89 blank (no signal) images (445 images in total). Blank images comprise 20\% of template signal dataset.

In our models we modify loss calculation described in basic IADB model which is simply a plain root mean square error (see Sec.~\ref{sec:IADB}). We incorporate SSIM metric into training process because we use the metric during similarity assessment between deblended and template images. Thus we multiply loss root mean square error by a weight factor, $w$, calculated using SSIM, in particular
\begin{equation}
w = \frac{1}{2}\Big(1 - \mathrm{SSIM}\big( D_{\theta}(x_{\alpha_t},\alpha_t), \, x_1 - x_0 \big)\Big) \, ,
\end{equation}
which ranges between $[0,1]$ and becomes $1$ in case of opposite image and $0$ in case of identical ones. Our loss is then
\begin{equation}
%l = \| w \cdot \big( D_{\theta}(x_{\alpha_t},\alpha_t) - (x_1 - x_0) \big) \|^2 \, .
l = w \cdot \| D_{\theta}(x_{\alpha_t},\alpha_t) - (x_1 - x_0) \|^2 \, .
\end{equation}

For training we use the Adam optimizer \cite{kingma2014adam}. We train the network in batches of single image, because all template images are blended with each noise image using random $\alpha$ coefficient value for each blending instance.

\begin{figure}
    \centering
        \centering
        \includegraphics[width=0.49\linewidth]{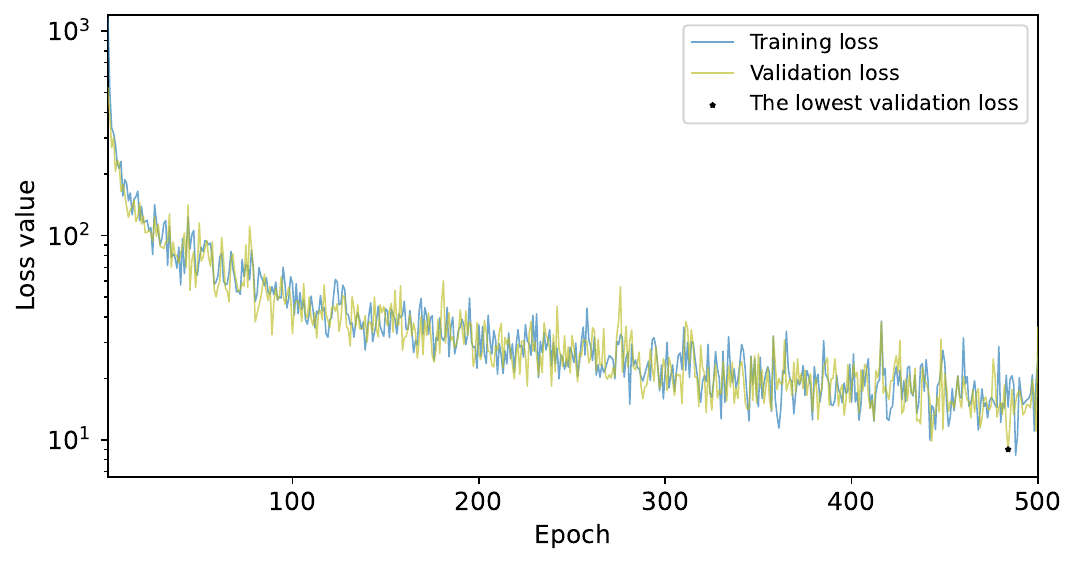}
        \includegraphics[width=0.49\linewidth]{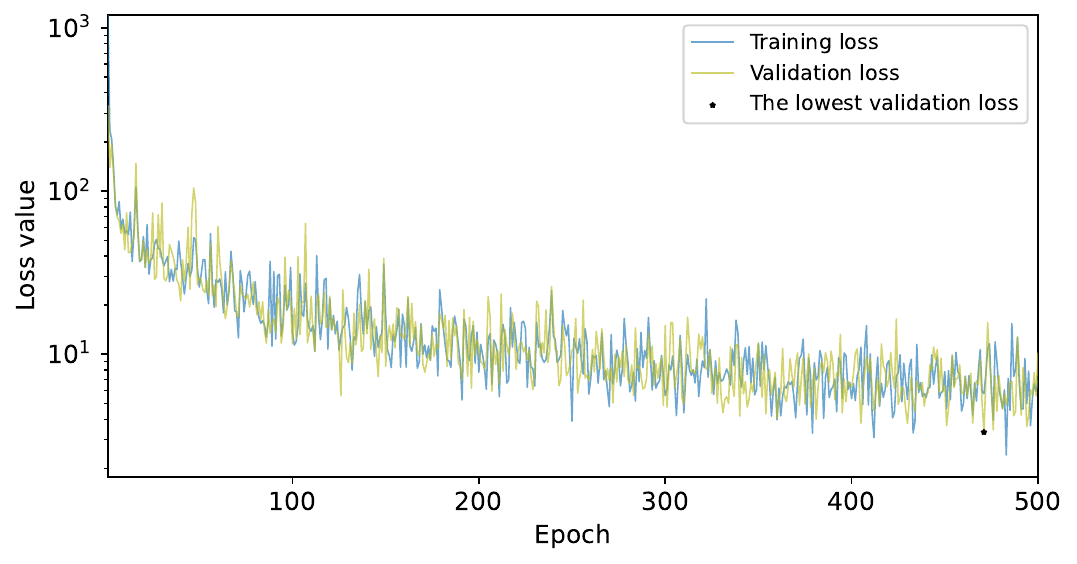}
    \caption{Training and validation loss evolution for Model~A (left panel) and Model~B (right panel).}
    \label{fig:LossEvolution}
\end{figure}

We train both models for 500 epochs until the training and validation loss functions stabilize (see Fig.~\ref{fig:LossEvolution}). For calculation of validation loss we use for each model a set of 6000 noise images not seen by the model during training. Then for inference we use trained model instances with the lowest validation loss value obtained which in our case are model instances at epoch 483 for Model~A and at epoch 470 for Model~B. Training takes about 2.5h at an nVidia V100 GPU.

\section*{References} 
\bibliography{references}

@ARTICLE{1998PhRvD..58f3001J,
       author = {Jaranowski, P. and Królak, A. and Schutz, B. F.},
        title = {{Data analysis of gravitational-wave signals from spinning neutron stars: The signal and its detection}},
      journal = {Phys Rev D},
         year = 1998,
        month = sep,
       volume = {58},
       number = {6},
          eid = {063001},
        pages = {063001},
          doi = {10.1103/PhysRevD.58.063001},
archivePrefix = {arXiv},
       eprint = {gr-qc/9804014},
 primaryClass = {gr-qc},
       adsurl = {https://ui.adsabs.harvard.edu/abs/1998PhRvD..58f3001J}
}

@ARTICLE{2015CQGra..32n5014P,
       author = {{Pisarski}, Andrzej and {Jaranowski}, Piotr},
        title = "{Banks of templates for all-sky narrow-band searches of gravitational waves from spinning neutron stars}",
      journal = {Classical and Quantum Gravity},
         year = 2015,
        month = jul,
       volume = {32},
       number = {14},
          eid = {145014},
        pages = {145014},
          doi = {10.1088/0264-9381/32/14/145014},
archivePrefix = {arXiv},
       eprint = {1302.0509},
 primaryClass = {gr-qc},
       adsurl = {https://ui.adsabs.harvard.edu/abs/2015CQGra..32n5014P}
}

@article{Pisarski_2023,
doi = {10.1088/1361-6382/acfe58},
url = {https://doi.org/10.1088/1361-6382/acfe58},
year = {2023},
month = {oct},
publisher = {IOP Publishing},
volume = {40},
number = {22},
pages = {225009},
author = {Pisarski, Andrzej and Jaranowski, Piotr},
title = {Banks of templates for directed and all-sky narrow-band searches of continuous gravitational waves from spinning neutron stars with several spindowns},
journal = {Classical and Quantum Gravity},
}

@inproceedings{alpha-deblending,
author = {Heitz, Eric and Belcour, Laurent and Chambon, Thomas},
title = {Iterative $\alpha$-(de)Blending: a Minimalist Deterministic Diffusion Model},
year = {2023},
isbn = {9798400701597},
publisher = {Association for Computing Machinery},
address = {New York, NY, USA},
url = {https://doi.org/10.1145/3588432.3591540},
doi = {10.1145/3588432.3591540},
booktitle = {ACM SIGGRAPH 2023 Conference Proceedings},
articleno = {34},
numpages = {8},
location = {Los Angeles, CA, USA},
series = {SIGGRAPH '23}
}

@misc{chen2024overview,
      title={An Overview of Diffusion Models: Applications, Guided Generation, Statistical Rates and Optimization}, 
      author={Minshuo Chen and Song Mei and Jianqing Fan and Mengdi Wang},
      year={2024},
      eprint={2404.07771},
      archivePrefix={arXiv},
      primaryClass={cs.LG},
      url={https://arxiv.org/abs/2404.07771}, 
}

@misc{yang2025,
      title={Diffusion Models: A Comprehensive Survey of Methods and Applications}, 
      author={Ling Yang and Zhilong Zhang and Yang Song and Shenda Hong and Runsheng Xu and Yue Zhao and Wentao Zhang and Bin Cui and Ming-Hsuan Yang},
      year={2025},
      eprint={2209.00796},
      archivePrefix={arXiv},
      primaryClass={cs.LG},
      url={https://arxiv.org/abs/2209.00796}, 
}

@misc{cao2023,
      title={A Survey on Generative Diffusion Model}, 
      author={Hanqun Cao and Cheng Tan and Zhangyang Gao and Yilun Xu and Guangyong Chen and Pheng-Ann Heng and Stan Z. Li},
      year={2023},
      eprint={2209.02646},
      archivePrefix={arXiv},
      primaryClass={cs.AI},
      url={https://arxiv.org/abs/2209.02646}, 
}

@ARTICLE{2020arXiv200611239H,
       author = {{Ho}, Jonathan and {Jain}, Ajay and {Abbeel}, Pieter},
        title = "{Denoising Diffusion Probabilistic Models}",
      journal = {arXiv e-prints},
         year = 2020,
        month = jun,
          eid = {arXiv:2006.11239},
        pages = {arXiv:2006.11239},
          doi = {10.48550/arXiv.2006.11239},
archivePrefix = {arXiv},
       eprint = {2006.11239},
 primaryClass = {cs.LG},
       adsurl = {https://ui.adsabs.harvard.edu/abs/2020arXiv200611239H}
}

@misc{LPIPS,
  title={The Unreasonable Effectiveness of Deep Features as a Perceptual Metric},
  author={Zhang, Richard and Isola, Phillip and Efros, Alexei A and Shechtman, Eli and Wang, Oliver},
  year={2018},
  howpublished = {CVPR, \url{https://github.com/richzhang/PerceptualSimilarity}},
}

@article{PhysRevD.95.062002,
  title = {Validating gravitational-wave detections: The Advanced LIGO hardware injection system},
  author = {Biwer, C. and others},
  journal = {\prd},
  volume = {95},
  issue = {6},
  pages = {062002},
  numpages = {15},
  year = {2017},
  month = {Mar},
  publisher = {American Physical Society},
  doi = {10.1103/PhysRevD.95.062002},
}

@MISC{GWOSC_CWHI, 
    author = "LVK",
    title = "O3 Hardware Injections", 
    howpublished = "{\tt https://gwosc.org/O3/o3\_inj}", 
    day = "10",
    month = "Nov", 
    year = "2025",
}

@MISC{tdfstat-repo, 
    author = "{Cieciel{\k a}g}, P. and others",
    collaboration = {Virgo-POLGRAW}, 
    title = "Time-domain F-statistic all-sky search pipeline", 
    howpublished = "{\tt https://github.com/Polgraw/TDFstat}", 
    day = "21",
    month = "Jun", 
    year = "2026",
}

@article{Aasi_2014,
doi = {10.1088/0264-9381/31/16/165014},
url = {https://doi.org/10.1088/0264-9381/31/16/165014},
year = {2014},
month = {aug},
publisher = {IOP Publishing},
volume = {31},
number = {16},
pages = {165014},
author = {Aasi, J and others},
title = {Implementation of an \mathcal{F}-statistic all-sky search for continuous gravitational waves in Virgo VSR1 data},
journal = {Classical and Quantum Gravity},
}

@article{PhysRevD.96.062002,
  title = {All-sky search for periodic gravitational waves in the O1 LIGO data},
  author = {Abbott, B. P. and others},
  collaboration = {LIGO Scientific Collaboration and Virgo Collaboration},
  journal = {\prd},
  volume = {96},
  issue = {6},
  pages = {062002},
  numpages = {35},
  year = {2017},
  month = {Sep},
  publisher = {American Physical Society},
  doi = {10.1103/PhysRevD.96.062002},
  url = {https://link.aps.org/doi/10.1103/PhysRevD.96.062002}
}

@article{PhysRevD.97.102003,
  title = {Full band all-sky search for periodic gravitational waves in the O1 LIGO data},
  author = {Abbott, B. P. and others},
  collaboration = {LIGO Scientific Collaboration and Virgo Collaboration},
  journal = {\prd},
  volume = {97},
  issue = {10},
  pages = {102003},
  numpages = {31},
  year = {2018},
  month = {May},
  publisher = {American Physical Society},
  doi = {10.1103/PhysRevD.97.102003},
  url = {https://link.aps.org/doi/10.1103/PhysRevD.97.102003}
}

@article{PhysRevD.100.024004,
  title = {All-sky search for continuous gravitational waves from isolated neutron stars using Advanced LIGO O2 data},
  author = {Abbott, B. P. and others},
  collaboration = {LIGO Scientific Collaboration and Virgo Collaboration},
  journal = {\prd},
  volume = {100},
  issue = {2},
  pages = {024004},
  numpages = {27},
  year = {2019},
  month = {Jul},
  publisher = {American Physical Society},
  doi = {10.1103/PhysRevD.100.024004},
  url = {https://link.aps.org/doi/10.1103/PhysRevD.100.024004}
}

@article{PhysRevD.106.102008,
  title = {All-sky search for continuous gravitational waves from isolated neutron stars using Advanced LIGO and Advanced Virgo O3 data},
  collaboration = {DLIGO Scientific Collaboration and Virgo Collaboration  and KAGRA Collaboration},
  author = {Abbott, R. and others},
  journal = {\prd},
  volume = {106},
  issue = {10},
  pages = {102008},
  numpages = {37},
  year = {2022},
  month = {Nov},
  publisher = {American Physical Society},
  doi = {10.1103/PhysRevD.106.102008},
  url = {https://link.aps.org/doi/10.1103/PhysRevD.106.102008}
}

@ARTICLE{1057571,
  author={Turin, G.},
  journal={IRE Transactions on Information Theory}, 
  title={An introduction to matched filters}, 
  year={1960},
  volume={6},
  number={3},
  pages={311-329},
  doi={10.1109/TIT.1960.1057571}}

@article{Cooley:1965zz,
    author = "Cooley, James W. and Tukey, John W.",
    title = "{An Algorithm for the Machine Calculation of Complex Fourier Series}",
    doi = "10.1090/S0025-5718-1965-0178586-1",
    journal = "Math. Comput.",
    volume = "19",
    pages = "297--301",
    year = "1965"
}

@article{Sieniawska2019,
    author = {Sieniawska, Magdalena and Bejger, Michał},
    title = {Continuous Gravitational Waves from Neutron Stars: Current Status and Prospects},
    doi = {10.3390/universe5110217},
    url = {https://www.mdpi.com/2218-1997/5/11/217},
    journal = {Universe},
    year = {2019},
    volume = {5},
    number = {11},
    article-number = {217},
    issn = {2218-1997}
}

@article{Riles2023,
    author={Riles, Keith},
    title={Searches for continuous-wave gravitational radiation},
    doi={10.1007/s41114-023-00044-3},
    url={https://doi.org/10.1007/s41114-023-00044-3},
    journal={Living Reviews in Relativity},
    year={2023},
    month={Apr},
    day={17},
    volume={26},
    number={1},
    issn={1433-8351},
}

@article{Haskell2023,
    author={Haskell, B. and Bejger, M.},
    title={Astrophysics with continuous gravitational waves},
    doi={10.1038/s41550-023-02059-w},
    url={https://doi.org/10.1038/s41550-023-02059-w},
    journal={Nature Astronomy},
    year={2023},
    month={Oct},
    day={01},
    volume={7},
    number={10},
    pages={1160-1170},
    issn={2397-3366},
}

@article{WETTE2023102880,
    author = {Karl Wette},
    title = {Searches for continuous gravitational waves from neutron stars: A twenty-year retrospective},
    doi = {https://doi.org/10.1016/j.astropartphys.2023.102880},
    url = {https://www.sciencedirect.com/science/article/pii/S092765052300066X},
    journal = {Astroparticle Physics},
    year = {2023},
    month={Nov},
    volume = {153},
    pages = {102880},
    issn = {0927-6505},
}

@article{PhysRevD.59.063003,
  title = {Data analysis of gravitational-wave signals from spinning neutron stars. II. Accuracy of estimation of parameters},
  author = {Jaranowski, Piotr and Kr\'olak, Andrzej},
  journal = {\prd},
  volume = {59},
  issue = {6},
  pages = {063003},
  numpages = {29},
  year = {1999},
  month = {Feb},
  publisher = {American Physical Society},
  doi = {10.1103/PhysRevD.59.063003},
  url = {https://link.aps.org/doi/10.1103/PhysRevD.59.063003}
}

@article{PhysRevD.61.062001,
  title = {Data analysis of gravitational-wave signals from spinning neutron stars. III. Detection statistics and computational requirements},
  author = {Jaranowski, Piotr and Kr\'olak, Andrzej},
  journal = {\prd},
  volume = {61},
  issue = {6},
  pages = {062001},
  numpages = {32},
  year = {2000},
  month = {Feb},
  publisher = {American Physical Society},
  doi = {10.1103/PhysRevD.61.062001},
  url = {https://link.aps.org/doi/10.1103/PhysRevD.61.062001}
}

@article{PhysRevD.65.042003,
  title = {Data analysis of gravitational-wave signals from spinning neutron stars. IV. An all-sky search},
  author = {Astone, Pia and Borkowski, Kazimierz M. and Jaranowski, Piotr and Kr\'olak, Andrzej},
  journal = {\prd},
  volume = {65},
  issue = {4},
  pages = {042003},
  numpages = {18},
  year = {2002},
  month = {Jan},
  publisher = {American Physical Society},
  doi = {10.1103/PhysRevD.65.042003},
  url = {https://link.aps.org/doi/10.1103/PhysRevD.65.042003}
}

@article{PhysRevD.82.022005,
  title = {Data analysis of gravitational-wave signals from spinning neutron stars. V. A narrow-band all-sky search},
  author = {Astone, Pia and Borkowski, Kazimierz M. and Jaranowski, Piotr and Pietka, Maciej and Kr\'olak, Andrzej},
  journal = {\prd},
  volume = {82},
  issue = {2},
  pages = {022005},
  numpages = {16},
  year = {2010},
  month = {Jul},
  publisher = {American Physical Society},
  doi = {10.1103/PhysRevD.82.022005},
  url = {https://link.aps.org/doi/10.1103/PhysRevD.82.022005}
}

@inbook{10.5555/3454287.3455354,
author = {Song, Yang and Ermon, Stefano},
title = {Generative modeling by estimating gradients of the data distribution},
year = {2019},
publisher = {Curran Associates Inc.},
address = {Red Hook, NY, USA},
booktitle = {Proceedings of the 33rd International Conference on Neural Information Processing Systems},
articleno = {1067},
numpages = {13},
chapter = {1067},
pages = {11918 - 11930}
}

@inproceedings{10.5555/3666122.3667767,
author = {Lim, Sungbin and Yoon, Eunbi and Byun, Taehyun and Kang, Taewon and Kim, Seungwoo and Lee, Kyungjae and Choi, Sungjoon},
title = {Score-based generative modeling through stochastic evolution equations in hilbert spaces},
year = {2023},
publisher = {Curran Associates Inc.},
address = {Red Hook, NY, USA},
booktitle = {Proceedings of the 37th International Conference on Neural Information Processing Systems},
articleno = {1645},
numpages = {14},
location = {New Orleans, LA, USA},
series = {NIPS '23}
}

@inproceedings{10.5555/3540261.3540933,
author = {Dhariwal, Prafulla and Nichol, Alex},
title = {Diffusion models beat GANs on image synthesis},
year = {2021},
isbn = {9781713845393},
publisher = {Curran Associates Inc.},
address = {Red Hook, NY, USA},
booktitle = {Proceedings of the 35th International Conference on Neural Information Processing Systems},
articleno = {672},
numpages = {15},
series = {NIPS '21}
}

@misc{ho2022classifierfreediffusionguidance,
      title={Classifier-Free Diffusion Guidance}, 
      author={Jonathan Ho and Tim Salimans},
      year={2022},
      eprint={2207.12598},
      archivePrefix={arXiv},
      primaryClass={cs.LG},
      url={https://arxiv.org/abs/2207.12598}, 
}

@inproceedings{10.5555/3540261.3541637,
author = {Austin, Jacob and Johnson, Daniel D. and Ho, Jonathan and Tarlow, Daniel and van den Berg, Rianne},
title = {Structured denoising diffusion models in discrete state-spaces},
year = {2021},
isbn = {9781713845393},
publisher = {Curran Associates Inc.},
address = {Red Hook, NY, USA},
booktitle = {Proceedings of the 35th International Conference on Neural Information Processing Systems},
articleno = {1376},
numpages = {13},
series = {NIPS '21}
}

@inproceedings{10.5555/3540261.3541214,
author = {Hoogeboom, Emiel and Nielsen, Didrik and Jaini, Priyank and Forr\'{e}, Patrick and Welling, Max},
title = {Argmax flows and multinomial diffusion: learning categorical distributions},
year = {2021},
isbn = {9781713845393},
publisher = {Curran Associates Inc.},
address = {Red Hook, NY, USA},
booktitle = {Proceedings of the 35th International Conference on Neural Information Processing Systems},
articleno = {953},
numpages = {12},
series = {NIPS '21}
}

@INPROCEEDINGS {9878449,
author = { Rombach, Robin and Blattmann, Andreas and Lorenz, Dominik and Esser, Patrick and Ommer, Bjorn },
booktitle = { 2022 IEEE/CVF Conference on Computer Vision and Pattern Recognition (CVPR) },
title = {{ High-Resolution Image Synthesis with Latent Diffusion Models }},
year = {2022},
volume = {},
ISSN = {},
pages = {10674-10685},
doi = {10.1109/CVPR52688.2022.01042},
url = {https://doi.ieeecomputersociety.org/10.1109/CVPR52688.2022.01042},
publisher = {IEEE Computer Society},
address = {Los Alamitos, CA, USA},
month =Jun}

@inproceedings{10.5555/3666122.3667911,
author = {Bansal, Arpit and Borgnia, Eitan and Chu, Hong-Min and Li, Jie S. and Kazemi, Hamid and Huang, Furong and Goldblum, Micah and Geiping, Jonas and Goldstein, Tom},
title = {Cold diffusion: inverting arbitrary image transforms without noise},
year = {2023},
publisher = {Curran Associates Inc.},
address = {Red Hook, NY, USA},
booktitle = {Proceedings of the 37th International Conference on Neural Information Processing Systems},
articleno = {1789},
numpages = {24},
location = {New Orleans, LA, USA},
series = {NIPS '23}
}

@inproceedings{10.5555/3618408.3619743,
author = {Song, Yang and Dhariwal, Prafulla and Chen, Mark and Sutskever, Ilya},
title = {Consistency models},
year = {2023},
publisher = {JMLR.org},
booktitle = {Proceedings of the 40th International Conference on Machine Learning},
articleno = {1335},
numpages = {42},
location = {Honolulu, Hawaii, USA},
series = {ICML'23}
}

@misc{lipman2023flowmatchinggenerativemodeling,
      title={Flow Matching for Generative Modeling}, 
      author={Yaron Lipman and Ricky T. Q. Chen and Heli Ben-Hamu and Maximilian Nickel and Matt Le},
      year={2023},
      eprint={2210.02747},
      archivePrefix={arXiv},
      primaryClass={cs.LG},
      url={https://arxiv.org/abs/2210.02747}, 
}

@INPROCEEDINGS{GarberTirer,
  author={Garber, Tomer and Tirer, Tom},
  booktitle={2024 IEEE/CVF Conference on Computer Vision and Pattern Recognition (CVPR)}, 
  title={Image Restoration by Denoising Diffusion Models with Iteratively Preconditioned Guidance}, 
  year={2024},
  volume={},
  number={},
  pages={25245-25254},
  doi={10.1109/CVPR52733.2024.02385}
}

@article{YaoLiZhichangYuming,
author = {Li, Yao and Cheng, Li and Guo, Zhichang and Xing, Yuming},
title = {Deep learning informed diffusion equation model for image denoising},
journal = {IET Image Processing},
volume = {18},
number = {13},
pages = {4310-4327},
doi = {https://doi.org/10.1049/ipr2.13253},
url = {https://ietresearch.onlinelibrary.wiley.com/doi/abs/10.1049/ipr2.13253},
eprint = {https://ietresearch.onlinelibrary.wiley.com/doi/pdf/10.1049/ipr2.13253},
year = {2024}
}

@ARTICLE{2015arXiv150303585S,
       author = {{Sohl-Dickstein}, Jascha and {Weiss}, Eric A. and {Maheswaranathan}, Niru and {Ganguli}, Surya},
        title = "{Deep Unsupervised Learning using Nonequilibrium Thermodynamics}",
      journal = {arXiv e-prints},
         year = 2015,
        month = mar,
          eid = {arXiv:1503.03585},
        pages = {arXiv:1503.03585},
          doi = {10.48550/arXiv.1503.03585},
archivePrefix = {arXiv},
       eprint = {1503.03585},
 primaryClass = {cs.LG},
       adsurl = {https://ui.adsabs.harvard.edu/abs/2015arXiv150303585S}
}

@misc{AlexNet,
      title={One weird trick for parallelizing convolutional neural networks}, 
      author={Alex Krizhevsky},
      year={2014},
      eprint={1404.5997},
      archivePrefix={arXiv},
      primaryClass={cs.NE},
      url={https://arxiv.org/abs/1404.5997}, 
}

@ARTICLE{SSIM_ref1,
  author={Zhou Wang and Bovik, A.C. and Sheikh, H.R. and Simoncelli, E.P.},
  journal={IEEE Transactions on Image Processing}, 
  title={Image quality assessment: from error visibility to structural similarity}, 
  year={2004},
  volume={13},
  number={4},
  pages={600-612},
  doi={10.1109/TIP.2003.819861}
}

@ARTICLE{SSIM_ref2,
  author={Avanaki, Alireza Nasiri},
  journal={Optical Review}, 
  title={Exact global histogram specification optimized for structural similarity}, 
  year={2009},
  volume={16},
  number={6},
  pages={613-621},
  doi={10.1007/s10043-009-0119-z}
}

@ARTICLE{ALIGO2015,
   author = {{Aasi}, J. and others},
    title = "{Advanced {LIGO}}",
  journal = {CQG},
     year = 2015,
   volume = 32,
   number = 7,
    pages = {074001},
      doi = {10.1088/0264-9381/32/7/074001},
   adsurl = {http://adsabs.harvard.edu/abs/2015CQGra..32g4001L}
}

@ARTICLE{AdV2015,
   author = {{Acernese}, F. and others},
    title = "{Advanced {Virgo}: a second-generation interferometric gravitational wave detector}",
  journal = {CQG},
     year = 2015,
   volume = 32,
   number = 2,
    pages = {024001},
      doi = {10.1088/0264-9381/32/2/024001},
   adsurl = {http://adsabs.harvard.edu/abs/2015CQGra..32b4001A}
}

@ARTICLE{KAGRA2013,
   author = {{Aso}, Y. and others},
    title = "{Interferometer design of the {KAGRA} gravitational wave detector}",
  journal = {PRD},
     year = 2013,
   volume = 88,
   number = 4,
    pages = {043007},
      doi = {10.1103/PhysRevD.88.043007},
   adsurl = {http://adsabs.harvard.edu/abs/2013PhRvD..88d3007A}
}

@ARTICLE{gwtc5,
       author = {{Abac}, A.~G. and others},
        title = "{GWTC-5.0: An Introduction to Version 5.0 of the Gravitational-Wave Transient Catalog}",
      journal = {arXiv e-prints},
         year = 2026,
        month = may,
          eid = {arXiv:2605.27223},
        pages = {arXiv:2605.27223},
          doi = {10.48550/arXiv.2605.27223},
archivePrefix = {arXiv},
       eprint = {2605.27223},
 primaryClass = {gr-qc},
       adsurl = {https://ui.adsabs.harvard.edu/abs/2026arXiv260527223T}
}

@article{GW150914,
       author = {{Abbott}, B.~P. and others},
        title = "{Observation of Gravitational Waves from a Binary Black Hole Merger}",
      journal = {\prl},
         year = 2016,
        month = feb,
       volume = {116},
       number = {6},
          eid = {061102},
        pages = {061102},
          doi = {10.1103/PhysRevLett.116.061102},
archivePrefix = {arXiv},
       eprint = {1602.03837},
 primaryClass = {gr-qc},
       adsurl = {https://ui.adsabs.harvard.edu/abs/2016PhRvL.116f1102A}
}

@ARTICLE{2025ApJ...983...99A,
       author = {{Abac}, A.~G. and others},
        title = "{Search for Continuous Gravitational Waves from Known Pulsars in the First Part of the Fourth LIGO-Virgo-KAGRA Observing Run}",
      journal = {\apj},
         year = 2025,
        month = apr,
       volume = {983},
       number = {2},
          eid = {99},
        pages = {99},
          doi = {10.3847/1538-4357/adb3a0},
archivePrefix = {arXiv},
       eprint = {2501.01495},
 primaryClass = {astro-ph.HE},
       adsurl = {https://ui.adsabs.harvard.edu/abs/2025ApJ...983...99A}
}

@ARTICLE{2026arXiv260325808T,
       author = {{Abac}, A.~G. and others},
        title = "{Searches for Continuous Gravitational Waves from Supernova Remnants in the first part of the LIGO-Virgo-KAGRA Fourth Observing run}",
      journal = {arXiv e-prints},
         year = 2026,
        month = mar,
          eid = {arXiv:2603.25808},
        pages = {arXiv:2603.25808},
          doi = {10.48550/arXiv.2603.25808},
archivePrefix = {arXiv},
       eprint = {2603.25808},
 primaryClass = {gr-qc},
       adsurl = {https://ui.adsabs.harvard.edu/abs/2026arXiv260325808T}
}

@ARTICLE{2026arXiv260314168T,
       author = {{Abac}, A.~G. and others},
        title = "{All-sky Searches for Continuous Gravitational Waves from Isolated Neutron Stars in the Data from the First Part of the Fourth LIGO-Virgo-KAGRA Observing Run}",
      journal = {arXiv e-prints},
         year = 2026,
        month = mar,
          eid = {arXiv:2603.14168},
        pages = {arXiv:2603.14168},
          doi = {10.48550/arXiv.2603.14168},
archivePrefix = {arXiv},
       eprint = {2603.14168},
 primaryClass = {gr-qc},
       adsurl = {https://ui.adsabs.harvard.edu/abs/2026arXiv260314168T}
}

@article{kingma2014adam,
  author = {Kingma, Diederik P and Ba, Jimmy},
  journal = {arXiv preprint arXiv:1412.6980},
  title = {Adam: A method for stochastic optimization},
  year = 2014
}

\end{document}